\documentclass[3p,times]{elsarticle}
\usepackage[english]{babel}     
\usepackage[utf8]{inputenc}   
\usepackage{amsmath}            
\usepackage{comment}
\usepackage{multirow}
\usepackage{xcolor}
\usepackage{float}
\begin{document}

\begin{frontmatter}

\title{Sapling Similarity: a performing and interpretable memory-based tool for recommendation}

\author[1,2]{Giambattista Albora\corref{cor1}}
\ead{alboragiambattista@gmail.com}
\author[1,3,5]{Lavinia Rossi Mori}
\author[4,1]{Andrea Zaccaria}

\cortext[cor1]{Corresponding Author.}

\address[1]{Enrico Fermi Research Center, Rome, Italy}
\address[2]{Physics department, La Sapienza University, Rome, Italy}
\address[3]{Physics department, Tor Vergata University, Rome, Italy}
\address[4]{Istituto dei Sistemi Complessi - CNR, UOS Sapienza, Rome, Italy}
\address[5]{Sony Computer Science Laboratories Rome, Joint Initiative CREF-Sony, Rome, Italy.}

\begin{abstract}
Many bipartite networks describe systems where an edge represents a relation between a user and an item. Measuring the similarity between either users or items is the basis of memory-based collaborative filtering, a widely used method to build a recommender system with the purpose of proposing items to users. When the edges of the network are unweighted, the popular common neighbors-based approaches, allowing only positive similarity values, neglect the possibility and the effect of two users (or two items) being very dissimilar. Moreover, they underperform with respect to model-based (machine learning) approaches, although providing a higher interpretability. Inspired by the functioning of Decision Trees, we propose a method to compute similarity that allows also negative values, the Sapling Similarity. The key idea is to look at how the information that a user is connected to an item influences our prior estimation of the probability that another user is connected to the same item: if it is reduced, then the similarity between the two users will be negative, otherwise it will be positive. We show that, when used to build memory-based collaborative filtering, Sapling Similarity provides better recommendations than existing similarity metrics. Then we compare the Sapling Similarity Collaborative Filtering (SSCF, an hybrid of the item-based and the user-based) with state-of-the-art models using standard datasets. Even if SSCF depends on only one straightforward hyperparameter, it has comparable or higher recommending accuracy, and outperforms all other models on the Amazon-Book dataset, while retaining the high explainability of memory-based approaches.
\end{abstract}

\end{frontmatter}

\section{Introduction}
Many complex systems can be reduced to the interaction between two classes of different objects; this is the case of several economic systems, where either countries or firms are connected to exported products or technological sectors \cite{albora2022machine, tacchella2012new, straccamore2021will}; biological systems, where either patients or microbes are connected with diseases \cite{goh2007human,li2019novel}; social systems, where for instance users are connected to Facebook pages \cite{del2017mapping,schmidt2018polarization,zaccaria2019poprank}, or actors are connected to movies they participate in \cite{watts1998collective}. An effective way to represent these systems is through bipartite networks, in which links can connect only nodes belonging to the two different sets. For instance, the bipartite network representing which country exports which product is the basis of the economic complexity (EC) framework \cite{hidalgo2021economic}. The main tool of EC is the so-called relatedness \cite{hidalgo2018principle,tacchella2021relatedness, albora2023product}, a measure of how much a country is close to start exporting a given product. This is a key tool for institutions and policymakers, and a driver for investments \cite{pugliese2020economic,lin2020african}. The traditional way of measuring the relatedness between a country $c$ and a product $p$ consists in analyzing the export basket of countries, extracting the similarity between $p$ and other products, and computing the average similarity between $p$ and the products exported by $c$ \cite{hidalgo2007product}. In the information system framework, this would be called an item-based Collaborative Filtering (CF) \cite{goldberg1992using, schafer2007collaborative}.\\
CF is one of the most popular methods to build a recommender system whose purpose is to suggest to users those items they will probably like. The idea is to base the recommendations on the previous interactions between users and items, collected in a user-item biadjacency matrix (that in the recommender system framework is usually called rating matrix). There are two main typologies of CF: memory-based and model-based \cite{chen2018survey}. The former obtains similarity measures between users or items according to the user-item adjacency matrix in order to find the nearest neighbors of a user or an item, then it bases the recommendations on these neighbors. The latter makes use of machine learning and data mining methods in order to achieve high recommendation quality. The advantages of a memory-based CF, with respect to a model-based one, include its simple and intuitive implementation, independence from hyper-parameters, and high interpretability of results \cite{ghazarian2015enhancing}. However, current state-of-the-art CF are model-based and use techniques such as Graph Convolution Networks (GCN) \cite{kipf2016semi} and Matrix Factorization (MF) \cite{koren2009matrix} that rely on optimized hyperparameters to maximize their performance. Despite this, due to their higher simplicity and interpretability, in practical applications such as the Economic Complexity framework, memory-based approaches are preferred to model-based ones and are largely used for both prediction and recommendation purposes \cite{hidalgo2007product, zaccaria2014taxonomy, pugliese2019unfolding}.\\
Talking about memory-based CF, there are two main approaches to give recommendations:
\begin{itemize}
    \item user-based: one measures the similarity between users based on how similar their links with items are and then suggests to a user an item that is popular among similar users;
    \item item-based: one measures the similarity between items and then suggests to a user an item that is similar to those to which it is already connected.
\end{itemize}
In both cases, when one measures the similarity between two nodes, there is a distinction between local and global similarity \cite{kumar2020link,papadakis2022collaborative}. The difference is that to compute a local similarity one needs information only about the two involved nodes, while to compute a global similarity also the whole graph structure is required; this makes the latter computationally demanding and hard to apply to large networks. When we deal with bipartite networks with unary (unweighted) links, which means that the corresponding biadjacency matrix element can be either equal to 1 if a link is present or 0 if it is absent, local similarities are usually based on co-occurrences, that is a count of how many times two nodes of the same layer are linked to the same node of the other layer. A key issue is that these metrics are positive-definite, and as a consequence, they neglect the information about the possible dissimilarity between two nodes. Xie et al \cite{xie2015link} observed that neglecting negative values in similarity metrics can be an issue when one deals with binary data (whose entries can be like (1), dislike (-1), or absence (0)); however, we argue that negative similarity should be taken into account also when we deal with unary data.\\
Let us make an example taken from economics. Allowing only positive similarity values does not take into account the fact that, for instance, Japan is specialized in high-tech products, while Zambia has a focus on simple products like raw materials. Knowing that Japan exports a product $p$ makes Zambia not exporting $p$ highly probable; on the other side, if Zambia exports $p'$, probably Japan is not interested in exporting it. So it makes sense to define a negative similarity between Zambia and Japan. In other words, we can say that Zambia and Japan are anti-correlated, so knowing that Zambia exports a product should have a negative effect on the recommendation of that product to Japan. It is therefore natural to distinguish the case in which two export baskets are independent (zero similarity) from a situation in which the fact that a country exports a product implies that another country does not (negative similarity).\\
In this paper, we propose a local similarity metric that allows also negative values, the Sapling Similarity. With a structure inspired by the functioning of decision trees \cite{quinlan1986induction}, Sapling Similarity is able to identify when two nodes are anti-correlated or dissimilar, in the sense that knowing that the first is connected to a node of the other layer, reduces the probability that also the second is connected to it. As we will better discuss later, the key to distinguishing anti-correlation from uncorrelation is to look at the maximum degree that a node can have (the total number of nodes in the other layer with respect to the one where the similarity is computed).  We will see that Sapling Similarity outperforms other similarities when used to build a memory-based CF for recommending items to users. Moreover, by combining the results provided by a user-based and an item-based CF based on Sapling Similarity, we define the SSCF (Sapling Similarity Collaborative Filtering). We also compare the performance of SSCF with the ones of current state-of-the-art model-based CF on three standard datasets commonly used to evaluate recommender systems \cite{wang2019neural,he2020lightgcn,choi2022perturbation}: Gowalla, Yelp2018, and Amazon-Book. Notably, on all dataset the performance of SSCF is comparable with model-based approaches; moreover, on the Amazon-Book dataset, it outperforms all existing models. This is a noteworthy result, since our memory-based CF is compared with machine learning-based methods whose hyperparameters are optimized on the test set. Moreover, SSCF is fully explainable, its only hyperparameter is not optimized on the test set and, not requiring a training phase, necessitates a lower computational effort.\\
The rest of the paper is organized as follows: Section 2 presents the main similarity metrics used in the literature to build memory-based CF and the current state-of-art model-based approaches. Section 3 introduces the Sapling Similarity and the SSCF giving a detailed explanation of its construction. In section 4 we describe our experimental setup, the datasets we use, the methods we compare with SSCF, and the performance indicators we adopt in this comparison. In section 5 we present the results dividing them into two parts: the first, in which we show that Sapling Similarity provides better recommendations than existing similarity metrics, and the second in which we compare SSCF with the current state-of-the-art. Section 6 concludes and proposes further developments.
\section{Related Works}
CF is a very popular technique used by recommender systems introduced in the mid-1990s \cite{adomavicius2005toward}. For instance, Amazon \cite{smith2017two}, for which recommendations are fundamental \cite{chen2007does}, uses item-based CF \cite{linden2003amazon} to recommend products to users. Also, in the Economic Complexity framework item-based CF is often used to measure relatedness, the affinity between a country and the export of a product \cite{hidalgo2007product,tacchella2021relatedness}.\\
CF works on bipartite networks where the nodes of one of the two layers can be identified as the users and the ones of the other layer as the items. Links between nodes represent rating users give to items; we can distinguish between four different cases according to the values that ratings can assume \cite{aggarwal2016recommender}:
\begin{itemize}
\item Continuous ratings: they can assume any value in an interval. For instance, in the Jester joke recommendation engine \cite{goldberg2001eigentaste}, ratings can take on any value between -10 and 10;
\item Interval-based ratings: they can assume a value from a discrete set of ordered numbers. For instance, Amazon and Netflix use a five-point rating scale, where ratings can be a value in the set \{1,2,3,4,5\};
\item Binary ratings: the user can give only a like or dislike for the item. For example, in social networks, the link between a user and a post can be -1 (the user dislikes the post), +1 (the user likes the post), or 0 (the user did not rate the post);
\item Unary ratings: the user can only rate positively an item, and there is no information about dislikes. So links can be either 0 if absent or 1 if present. For instance, a country may export a product (link present) or it may not export it (link absent).
\end{itemize}
In this paper, we focus on unary ratings.
\subsection{Memory-based CF}
The key point of a memory-based CF is the measure of the similarity between two users or items. When dealing with unary data, the simplest way to measure it is to count the number of co-occurrences, that is how many common neighbors the two users or items have \cite{liben2007link}. However, since nodes with a high degree tend to have more co-occurrences, so a higher similarity with other nodes, than nodes with a low degree \cite{cimini2022meta}. For this reason, usually, the degrees of the two nodes on which we measure the similarity are used to normalize the number of co-occurrences. Depending on this normalization factor, different metrics can be defined. In this paper we will consider: Jaccard Similarity \cite{jaccard1901etude}, Cosine Similarity \cite{salton1983introduction}, Sorensen Index \cite{sorensen1948method}, Hub Depressed Index, and Hub Promoted Index \cite{ravasz2002hierarchical}.\\
One can also weigh differently each co-occurrence, for instance, if we want to measure the similarity between two products and they are both exported by China, this is not a strong signal that they are similar, since China exports many products that can be very different one from each other. If instead, we know that the two products are both exported by Angola, they are probably similar because Angola exports only a few simple products. So a co-occurrence in Angola should weigh more than a co-occurrence in China, and for this reason, we can weigh each co-occurrence by dividing it by the degree of the node on which there is the co-occurrence or with the logarithm of the degree. In the former case we have the Resource Allocation index \cite{zhou2009predicting}, and in the latter one we have the Adamic/Adar index \cite{adamic2003friends}.\\
Using both the approaches we just described one can build other metrics, for instance Zhou et al. introduced the Probabilistic Spreading \cite{zhou2007bipartite}, and in the Economic Complexity framework Zaccaria et al. introduced the Taxonomy Network \cite{zaccaria2014taxonomy} that is the Hub Depressed Index with co-occurrences weighed in the Resource Allocation way.\\
The above-mentioned similarity metrics represent a key tool when investigating recommender systems with unary (weightless) ratings \cite{aggarwal2016recommender,papadakis2022collaborative}. It is important to note that such metrics are all positive-definite, which means that the similarity between two nodes can not be negative. Let us consider an item-based CF, which measures the similarity between items and recommends to a user an item that is similar to the ones already connected to that user. A positive-definite similarity hides the possibility that the connection between a user $u$ and an item $i$ gives a negative contribution to the recommendation of another item $j$. Let us suppose that $i$ and $j$ are anti-correlated, in the sense that usually users that are connected to $i$ are not connected to $j$. In this case, we would like to be able to assess how much the information that $u$ is connected to $i$ lowers our confidence about the fact that $u$ is also connected to $j$. The main limit of the similarity metrics based on co-occurrences we described above is their inability to assess and use the possible anti-correlation between two nodes.\\
When dealing with continuous or interval-based ratings, the anti-correlation between nodes can be expressed by the Pearson Similarity \cite{shardanand1995social}. For instance, if two users give opposite ratings to items (when the first user gives the maximum rate to an item, the second user gives the minimum one), then they are anti-correlated and their Pearson Similarity is -1. To the best of our knowledge, books and reviews on Collaborative Filterings \cite{aggarwal2016recommender,papadakis2022collaborative} do not report Pearson Similarity as a metric for unary data, because in this case, ratings can only be present or absent, so users can not give anti-correlated ratings to an item. However, in our study, we will also consider Pearson Similarity to provide a more complete comparison, by looking at the correlation between given and ungiven ratings.\\ 
Our proposed metric, the Sapling Similarity, is designed to work with unary ratings assigning positive values to nodes that are similar and negative values to the ones that are dissimilar and, with respect to the other metrics based on co-occurrences, takes explicitly into account the size of the network. 
\subsection{Model-based CF}
With the great success of machine learning techniques in different fields of research, more and more model-based CF techniques are being developed \cite{papadakis2022collaborative}. In particular, Graph Convolution Networks (GCN) \cite{chen2020revisiting} show high performance when used to build a CF. In 2019 NGCF (Neural Graph Collaborative Filtering) was introduced \cite{wang2019neural} and it achieved the best performance on the three datasets Gowalla, Yelp2018, and Amazon-Books. In 2020 He et al. with their proposed LightGCN \cite{he2020lightgcn} showed that by simplifying the architecture of NGCF one can achieve better recommendation quality improving the results on all three datasets. From LightGCN other GCNs were built, aiming to achieve optimal results with simple architectures: LT-OCF \cite{choi2021lt}, SimpleX \cite{mao2021simplex}, and UltraGCN \cite{mao2021ultragcn}. Finally, Choi et al. in 2022, inspired by score-based generative models \cite{song2020score}, introduced the blurring-sharpening process models BSPM-EM and BSPM-LM \cite{choi2022perturbation} that currently achieve the state-of-the-art results with all the three datasets\footnote{https://paperswithcode.com/task/recommendation-systems}.
All the above-mentioned model-based approaches depend on a number of hyperparameters that are optimized in order to obtain maximal recommendation scores on each dataset.
\section{Methodology}
In this section we introduce the Sapling Similarity, illustrating its formulation and an intuitive explanation of its functioning;  and the SSCF, a collaborative filtering based on this similarity measure. We start by providing the basic definitions needed and by fixing the notation. 
\subsection{Basic definitions}
A \textbf{bipartite network} is defined as a graph $G=(U,\Gamma,E)$ where $U$ and $\Gamma$ are two sets of nodes (called also layers), and E is the set of all the connections $(i,\alpha)$ between the nodes $i\in U$ and $\alpha\in \Gamma$. Let $|U|$ be the dimension (cardinality) of the set $U$ and $|\Gamma|$ the dimension of the set $\Gamma$: the bipartite network can be represented by a $|U|\times|\Gamma|$ binary matrix $M$ called \textbf{bi-adjacency matrix} and defined as:
\begin{equation}
	M_{i\alpha} =
	\begin{cases}
		1 ~~~\text{ if } (i,\alpha) \in E\\[3mm]
		0 ~~~\text{ if } (i,\alpha) \notin E
	\end{cases}	
\end{equation}
The \textbf{degree} of a node $i\in U$ or $\alpha\in \Gamma$ is the number of links connected to it:
\begin{equation}
    k_i=\sum_{\lambda=1}^{|\Gamma|}M_{i\lambda}~~~~~k_\alpha=\sum_{l=1}^{|U|}M_{l\alpha}
\end{equation}
The number of \textbf{co-occurrences} between either nodes $i,j\in U$ or $\alpha,\beta\in \Gamma$ is:
\begin{equation}
    CO^{(users)}_{ij}=\sum_{\lambda=1}^{|\Gamma|}M_{i\lambda}M_{j\lambda}~~~~~CO^{(items)}_{\alpha\beta}=\sum_{l=1}^{|U|}M_{l\alpha}M_{l\beta}
\end{equation}
Note that the CO matrices define two monopartite networks whose nodes belong to only one set.\\
Usually, a bipartite network describes interactions between users and items (for instance, in the case of the country-exported product network we can identify the former as the user and the latter as the item). In our notation, the set U corresponds to users and the set $\Gamma$ corresponds to items.\\
In the following, we will focus on the Sapling Similarity between two users, and we will define $N = |\Gamma|$ (the total number of items). The case of the similarity between items is equivalent, with the only changes that $N = |U|$ (the total number of users) and the co-occurrences and degrees are referred to the nodes in the layer of the items.
\subsection{The Decision Sapling}
\begin{figure*}[h!]
	\centering
		\includegraphics[width=0.99\textwidth]{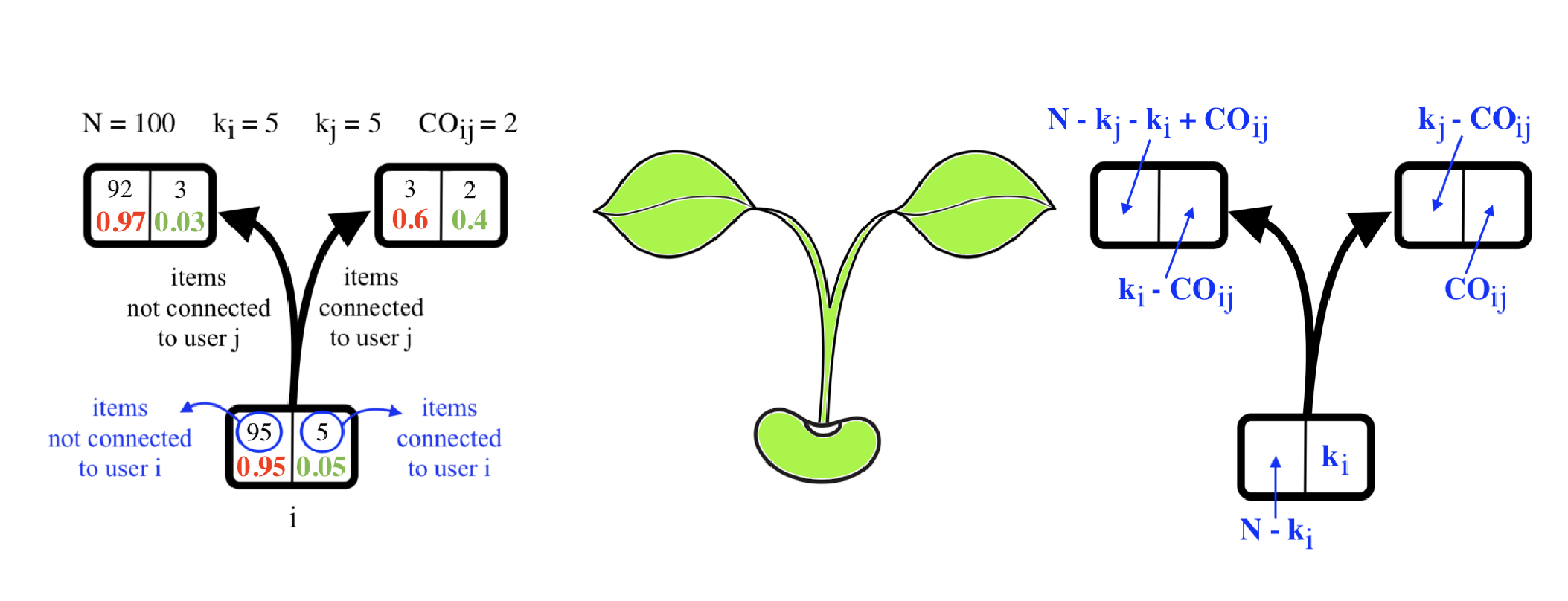}
		\caption{The Decision Sapling of user $i$ with respect to user $j$ is a tool to visualize and quantify how the probability that a generic item is connected to $i$ changes when one considers also the connections of $j$. On the left there is a numerical example and on the right the generic formulations. In the bean box at the base, we have two areas: on the right there is the number and the fraction of items user $i$ is connected to, while on the left the items user $i$ is not connected to. On the right leaf the same numbers are computed by restricting the number of items to the ones user $j$ is connected. Finally, on the left leaf only the items to which user $j$ is not connected are considered. By comparing the fractions in the bean with the ones in the leaves one can deduce whether the similarity between $i$ and $j$ is positive or negative.}
		\label{fig:little_tree}
\end{figure*}

The building block of the Sapling Similarity between two users $i$ and $j$ is what we call the Decision Sapling, which is a decision tree with just one split. The Decision Sapling is a diagram that represents and quantifies how much the information that user $j$ is or is not connected to an item $\alpha$ influences our estimate of the probability that another user $i$ is connected to $\alpha$. In figure \ref{fig:little_tree}, on the left, we show a numerical example where we build the Decision Sapling of user $i$ with respect to user $j$; on the right, we show the formulation of the generic case in terms of $CO^{(users)}_{ij}$, $k_i$, $k_j$, and $N$.
A Decision Sapling is composed of three boxes and each box is divided into two areas. In the right (left) area of all boxes there are the total number and the fraction of items to which user $i$ is (is not) connected. In the lower box (the bean) these numbers are computed with respect to all N items: so on the right we have $k_i$ (the number of items user $i$ is connected to) and on the left we have $N-k_i$. In the right box (right leaf) the numbers are computed by considering only the subset of items to which user $j$ is connected: so on the right area we have the co-occurrences $CO^{(users)}_{ij}$ (the number of items connected to both $i$ and $j$) and on the left area we will have the number of items connected to $j$ and not to $i$. Note that the sum is equal to $k_j$. In the left box (left leaf) we consider instead only the items not connected to user $j$; so for instance in the left box we will have the number and fraction of items not connected to $i$ nor to $j$. Note that the fractions are always computed with respect to the total number of items in the box, so the denominator is $N$ for the bean, $k_j$ for the right box, and $N-k_j$ for the left box. Let us now discuss a numerical example, reported on the left side of figure \ref{fig:little_tree}. We have a total of $N=100$ items; users $i$ and $j$ are connected to only $5$ of them, with $2$ common items. The a priori probability for an item to be connected with $i$ is provided in the bean and it is equal to 5\%. However, if we then look at the right leaf it emerges that by selecting the subset of items connected to $j$ the probability that also $i$ is connected to it varies from 5\% to 40\%. So
knowing that an item is connected to $j$ increases the probability that the same item is also connected to $i$: in this case, the similarity between $i$ and $j$ has to be positive.  Looking at the left leaf we see that knowing that $j$ is not connected to an item decreases the probability that $i$ is connected to it, so also from this point of view it is natural to give a positive similarity between $i$ and $j$ (the variation of the probabilities with respect to the bean is more evident in the right leaf because of the lower number of items connected to $j$). \\
We can modify this example to understand why, even before looking at the results of the prediction exercise, Sapling Similarity uses more features of the other similarity metrics, so providing a more comprehensive view. To the best of our knowledge, existing similarity metrics use the numerical values of $CO^{(users)}_{ij}$, $k_i$, and $k_j$, but not the total number of items $N$. Let us consider $N=8$ instead of $100$ in the previous example, and we leave the other numbers unchanged. In this case, the bean shows that $i$ is connected to 5 out of 8 items, which means 62.5\% of them, instead of the 5\% of the previous case. If we consider the right leaf, that is the subset of items connected to $j$, the numbers do not change and so $i$ is still connected to the 40\%. In this case, knowing that $j$ is connected to an item reduces the probability that also $i$ is connected to it. The similarity between $i$ and $j$ has to be negative. In conclusion, by only varying $N$ the similarity changes sign. Existing similarity metrics not only disregard the possibility of negative similarities, but also do not consider the information coming from the total dimension of the sets, since $N$ does not enter the equations.

\subsection{The Sapling Similarity: key idea}
\begin{figure*}[h!]
	\centering
		\includegraphics[width=0.9\textwidth]{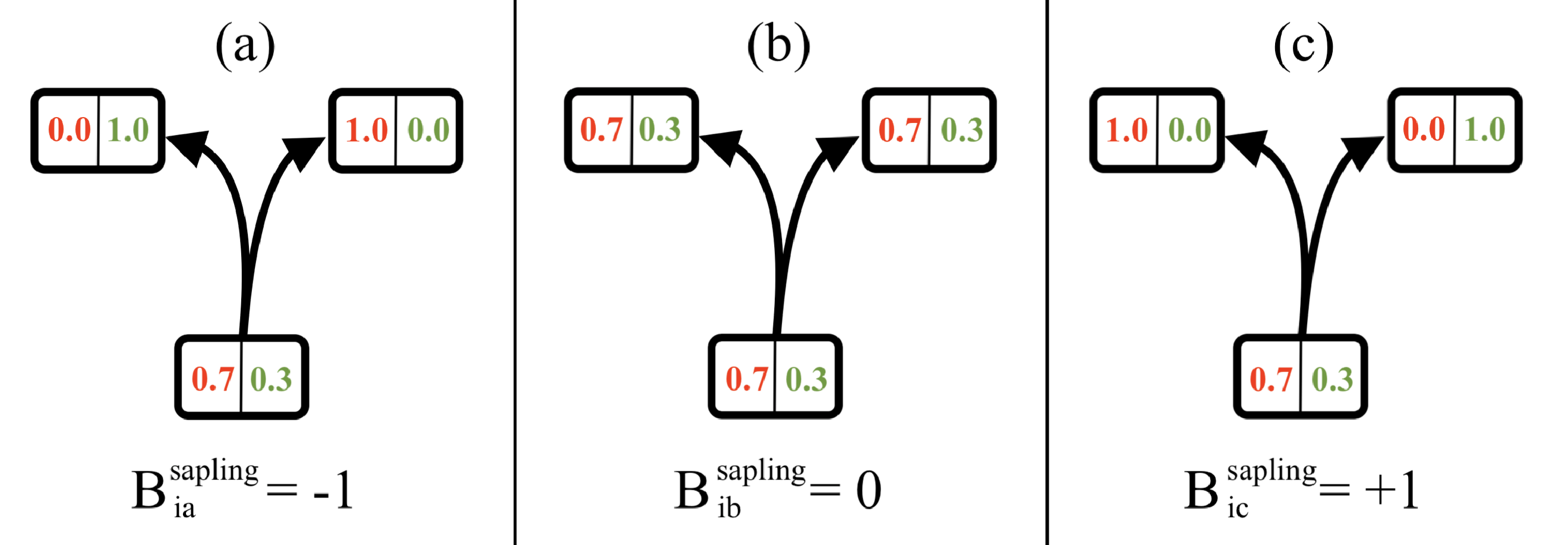}
		\caption{Three examples of Decision Saplings when computing the similarity between user $i$ and three users $a$, $b$, and $c$. (a) knowing the user $a$ is connected to an item $\alpha$ gives us the certainty that user $i$ is not connected to it. In this case, we say that user $i$ and $a$ are different and their Sapling Similarity is -1. (b) knowing the user $b$ is connected to an item $\alpha$ does not modify our prior knowledge that also $i$ is connected to it. In this case, we say that the Sapling Similarity between user $i$ and $b$ is 0. (c) knowing the user $c$ is connected to an item $\alpha$ gives us the certainty that also user $i$ is connected to it. In this case, we say that the Sapling Similarity between user $i$ and $c$ is +1.}
		\label{fig:boxes}
\end{figure*}

In this section, we discuss how the Decision Sapling tool can be used to determine the sign and the magnitude of the similarity between two items. In figure \ref{fig:boxes} we show three examples of Decision Saplings, namely of the same user $i$ with respect to three different users $a$, $b$, and $c$.
In the case (a) the Sapling Similarity $B_{ia}^{sapling} = -1$ since the knowledge that $a$ is linked to an item implies that $i$ is not linked to it: in particular, $a$ and $i$ have zero co-occurrences. In case (b) $B_{ib}^{sapling} = 0$ since the fact that $b$ is linked to $\alpha$ does not add information about the possibility that also $i$ is connected to $\alpha$. In case (c) $B_{ic}^{sapling} = 1$ since if one knows that $c$ is linked to $\alpha$, then for sure also $i$ is linked to it. In particular, here all $j$'s items co-occur with $i$. Looking at the left leaf it is easy to reach the same conclusion: for instance, knowing that $a$ is not connected to a node gives certainty that $i$ is connected to it, so $B_{ia}^{sapling} = -1$. In the cases in which the information provided by the items of $j$ removes all the uncertainty present in the bean, the Sapling Similarity is maximally positive or negative. If considering the subset of items connected to $j$ does not add any information about $i$, the similarity is zero. In order to provide an assessment of the similarity in the intermediate cases we make use of the Gini Impurity (GI). For each box we define:
\begin{equation}
GI = 1-p_1^2-p_0^2 = 2p_1p_0
\end{equation}
where $p_0$ is the fraction of zeros (the red values in figure \ref{fig:boxes}) and $p_1$ is the fraction of ones (the green values in figure \ref{fig:boxes}). The Gini Impurity reaches its minimum value of 0 when $p_0$ or $p_1$ is 0 and it reaches its maximum of 0.5 when $p_0 = p_1 = 0.5$. So the lower it is the GI the higher the polarization or certainty associated with the box. The relative variation of the Gini Impurity after the split given by user $j$ is:
\begin{equation}
\Delta GI = \frac{GI^{(b)}-f^{(l)}GI^{(l)}-f^{(r)}GI^{(r)}}{GI^{(b)}}
\label{eq:gini_variation}
\end{equation}
where $GI^{(b)}$ is the Gini Impurity of the bean, $GI^{(l)}$ the GI of the left leaf and $GI^{(r)}$ the GI of the right leaf, while $f^{(l)}$ is the fraction of total elements in the left leaf $\frac{N-k_{j}}{N}$, and $f^{(r)}$ the same fraction in the right leaf $\frac{k_{j}}{N}$. Notice that we normalize the Gini variation with $GI^{(b)}$ in order to have a $\Delta GI$ ranging between 0 and 1. $\Delta GI$ close to 0 means that considering $j$ does not vary the fractions in the leaves with respect to the bean, while $\Delta GI$ close to 1 means that the fractions after the split are more polarized than the ones in the bean, and so reduce the starting uncertainty. $\Delta GI$ is the absolute value of our similarity measure, the sign will be positive if the percentage of positive elements on the right leaf is higher than the one on the bean and negative otherwise.\\
Looking at the three examples in figure \ref{fig:boxes}, in the (a) and (c) cases the Gini Impurity of the leaves is 0, so according to equation \ref{eq:gini_variation}, $\Delta GI = \frac{GI^{(b)}}{GI^{(b)}} = 1$. Since in case (a) the percentage of positive elements on the right leaf is 0, so lower than the one on the bean, $B_{ia}^{sapling} = -1$, while since in case (c) the percentage of positive elements on the right leaf is 100\%, so higher than the one on the bean (30\%), $B_{ia}^{sapling} = +1$. In the case (b) the Gini Impurity in the left and right leaves is the same as the one in the bean, since $f^{(l)}+f^{(r)} = 1$ and $GI^{(b)} = GI^{(r)} = GI^{(l)}$ we will have $B_{ib}^{sapling} = 0$.
\subsection{The Sapling Similarity: formula}
Here we finally derive the Sapling Similarity formula. In particular, we express equation \ref{eq:gini_variation} in terms of the values that can be directly computed from the network, that is the total number of items N, the degrees of user $i$ ($k_i$) and $j$ ($k_j$), and the number of co-occurrences between $i$ and $j$ ($CO^{(users)}_{ij}$). We recall that each box is divided into a right area (which considers the items connected to $i$) and a left area (items not connected to $i$).
\begin{itemize}
\item \textbf{Bean:} The bean considers a total of $N$ elements, $k_{i}$ of which are connected to $i$; so $N-k_{i}$ items are not connected to $i$.
\item \textbf{Right leaf:} The total number of elements considered in the right box is $k_{j}$. Of those, $CO^{(users)}_{ij}$ are connected to $i$, and the number of items connected to $j$ but not to $i$ is $k_{j}-CO^{(users)}_{ij}$.
\item \textbf{Left leaf:} Here the total number is $N-k_{j}$. The number of elements connected to $i$ but not to $j$ is $k_{i}-CO^{(users)}_{ij}$. $N-k_{j}-k_{i}+CO^{(users)}_{ij}$ items are not connected to $i$ nor to $j$.
\end{itemize}
So the terms in equation \ref{eq:gini_variation} can be written as:\\
\begin{align*}
    &GI^{(b)}=2\frac{k_{i}}{N}\frac{N-k_{i}}{N}\\\
    &f^{(r)}=\frac{k_{j}}{N}\\
    &f^{(l)}=\frac{N-k_{j}}{N}\\
    &GI^{(r)}=2\frac{CO^{(users)}_{ij}}{k_{j}}\frac{k_{j}-CO^{(users)}_{ij}}{k_{j}}\\
    &GI^{(l)}=2\frac{k_{i}-CO^{(users)}_{ij}}{N-k_{j}}\frac{N-k_{j}-k_{i}+CO^{(users)}_{ij}}{N-k_{j}}
\end{align*}
and with a simple computation we can show that equation \ref{eq:gini_variation} is equivalent to:
\begin{equation}
\Delta GI = 1-f_{ij}
\end{equation}
where:
\begin{equation}
f_{ij} = \frac{CO_{ij}(1-\frac{CO_{ij}}{k_{j}})+(k_{i}-CO_{ij})(1-\frac{k_{i}-CO_{ij}}{N-k_{j}})}{k_{i}(1-\frac{k_{i}}{N})}.
\label{eq:tree_sim}
\end{equation}
In equation \ref{eq:tree_sim} we omitted the superscript (users) over the CO matrix since the formula is valid also when one measures the similarity between items. If $i$ and $j$ are users we will use $CO^{(users)}_{ij}$ and N will be the number of items, otherwise, if $i$ and $j$ are items, we will use $CO^{(items)}_{ij}$ and N will be the number of users.\\
As already said, $\Delta GI$ is the absolute value of our similarity measure. The sign of the Sapling Similarity is positive if the fraction of elements in the right area of the right leaf is bigger than the one on the bean, that is if $CO_{ij}/k_{j}\geq k_{i}/N$. If $CO_{ij}/k_{j}< k_{i}/N$, the sign of the Sapling Similarity is negative.\\

\noindent
\textbf{Observation.} \textit{Notice that $\Delta GI$ is singular if $k_{i} = 0$, $k_{i} = N$, $k_{j} = 0$ or $k_{j} = N$. If a node $l$ has degree $k_{l} = 0$ then it is not connected to anything, so it is useless for our collaborative filtering and it can be deleted from the network. If $k_{l} = N$ then $l$ is connected to all items; also this case the node does not bring useful information for our collaborative filtering.}\\

\noindent
In conclusion, the Sapling Similarity metric can be written as:
\begin{equation}
	B_{ij}^{sapling} =
	\begin{cases}
		1-f_{ij} ~~~\text{ if } \frac{CO_{ij}N}{k_{i}k_{j}}\geq1\\[3mm]
		-1+f_{ij} ~~~\text{ otherwise }
	\end{cases}	
\label{eq:sapling}
\end{equation} 
\noindent
\textbf{Lemma.} \textit{$f_{ij}$ is symmetric.}\\
\textbf{Proof.} \textit{we can write equation \ref{eq:tree_sim} as following:}
\begin{equation}
f_{ij} = N\frac{k_ik_j(2CO_{ij}-k_i-k_j)+N(k_ik_j-CO_{ij}^2)}{k_ik_j(N-k_i)(N-k_j)}
\end{equation}
\textit{it is trivial to prove that switching the indices i and j the result does not change (note that $CO_{ij}$ is symmetric).}\\
\textbf{Observation.} \textit{if $f_{ij}$ is symmetric then also  $B_{ij}^{sapling}$ is symmetric.}
\subsection{Sapling Similarity Collaborative Filtering}
After the assessment of the Sapling Similarity matrices $B^{(user)}$ and $B^{(item)}$, we can use them to build a user-based and an item-based CF. The confidence values for the recommendation of item $\alpha$ to user $i$ are given by the following equations \cite{sarwar2001item}:
\begin{equation}
S^{(user)}_{i\alpha}=\frac{\sum_{l}B^{(user)}_{il}M_{l\alpha}}{\sum_{l}|B^{(user)}_{il}|}~~~\text{ (user based)}
	\label{eq:density_user}
\end{equation}
\begin{equation}
S^{(item)}_{i\alpha}=\frac{\sum_{\lambda}B^{(item)}_{\alpha\lambda}M_{i\lambda}}{\sum_{\lambda}|B^{(item)}_{\alpha\lambda}|}~~~\text{ (item based)}
	\label{eq:density_item}
\end{equation}
We define the SSCF as a weighted average of the item-based and the user-based estimations:
\begin{equation}
	\text{SSCF}= (1-\gamma)S^{(user)}+\gamma S^{(item)}
	\label{eq:hybrid}
\end{equation}
$\gamma$ is the only parameter in our model and it regulates the relative weight we give to the user-based and the item-based recommendations in this \textit{hybrid} prescription.

\section{Experimental setup}
In this section, we describe the experimental setup we adopt to evaluate the performance of our proposed SSCF model and compare it with the ones proposed in the literature.
\subsection{Datasets}
In this study, we make use of 6 datasets:
\begin{itemize}
	\item Country-export dataset: the countries are interpreted as users and the exported products as items. The element of the M matrix is 1 if the Revealed Comparative Advantage (RCA) \cite{balassa1965trade} is greater than 1 (the country competitively exports the product) and 0 otherwise. The data about the export volumes of countries are provided by the UN-COMTRADE dataset (comtrade.un.org).
	\item Amazon-Product dataset: the users are Amazon accounts and the items the products they bought and rated from 1 to 5 \cite{mcauley2015image,pasricha2018translation}. The element of the M matrix is 1 if the user gave a rate of at least 3 to the item;
	\item Milan GPS data: users are inhabitants and the items are hexagonal areas \cite{biazzo2019general} containing Points of Interest (POI) in Milan extracted from OpenStreetMap (https://www.openstreetmap.org/). Data are provided by Cuebiq Inc. (https://www.cuebiq.com/about/data-for-good/). The element of the M matrix is 1 if the inhabitant visited at least one POI in the hexagonal area;
	\item Gowalla dataset: the users are individuals who use the Gowalla social network and perform check-ins, while the items are the locations or places where the check-ins are performed. \cite{he2020lightgcn};
	\item Yelp2018 dataset: the users are individuals who created an account on the Yelp platform, the items are businesses that have been reviewed by users \cite{he2020lightgcn};
	\item Amazon-Book dataset: the users are individuals who provide reviews for books on the Amazon platform, while the items are the books that are being reviewed \cite{he2020lightgcn};
\end{itemize}
Some properties of the datasets are shown in table \ref{tab:dataset}. The heterogeneity of these datasets allow for a fair and complete comparison among the different CFs. Further information regarding the first three datasets can be found in the supplementary material. The Gowalla, Yelp2018, and Amazon-Book datasets are commonly employed for benchmarking collaborative filtering techniques and are well-documented in the literature \cite{he2020lightgcn, choi2021lt,mao2021ultragcn,mao2021simplex,choi2022perturbation}. We refer interested readers to these sources for a more detailed description.
\begin{table*}[h!]
	\centering
	\begin{tabular}{l|cccc}
		dataset & \# users & \# items & \# interactions & Density \\
		\hline
		Country-Export & 169 & 5,040 & 120,438 & 14.14\%\\
		Amazon-product & 6,121 & 2,744 & 172,206 & 1.025\%\\
		Milan GPS data & 2,783 & 1,419 & 41,852 & 1.060\%\\
		Gowalla & 29,858 & 40,981 & 1,027,370 & 0.084\%\\
		Yelp2018 & 31,668 & 38,048 & 1,561,406 & 0.130\%\\
		Amazon-Book & 52,643 & 91,599 & 2,984,108 & 0.062\%
	\end{tabular}
	\caption{Main properties of our four datasets.}
	\label{tab:dataset}
\end{table*}

\subsection{Train-Test split}
Each dataset is divided in a train and a test set. The former is used to build the CF and the latter is used to evaluate the accuracy of the recommendations.\\
The train set in the country-export dataset is built from data on the export volume of countries between 1996 and 2013. The test set is composed of products that countries did not export in 2013 and previous years but exported in 2018.\\
The train in the Amazon-product dataset is the whole matrix M, but from each user, the last rated product in chronological order has been removed and it has been used for the test.\\
The data in the Milan GPS dataset cover a period of 9 months. We used the first 6 months for the train and the last 3 months for the test (we deleted links already present in the train set).\\
Finally, for the last three datasets, the train-test split is the same used by other papers in the literature \cite{he2020lightgcn,choi2021lt,mao2021simplex,mao2021ultragcn,choi2022perturbation} in order to guarantee a fair comparison with other Collaborative Filtering models.\\
The model-based methods we will consider in section \ref{sec:model-based} rely on hyperparameters whose value influences the recommendations. To set these values, the papers in the literature look at those maximizing the performance in the test set. However, in a real-world scenario test data are unknown, so they can not be used to optimize the hyperparameters. For this reason, in order to optimize $\gamma$ (the only parameter in our SSCF model) from the train data, we defined a validation set by removing from each user the 10\% of items, rounded up. We measured $B^{(user)}$ and $B^{(item)}$ using the remaining train data and we looked at the performance on the validation set.\\
Once we found the optimal value we used it to build the SSCF model using all train data. So the results we will show for SSCF do not depend on any hyperparameter that has been optimized on the test set; this is an important difference with respect to the current state-of-the-art models. 

\subsection{Other memory-based Collaborative Filtering}\label{sec:similarities}
As already stated, the first step to building a memory-based Collaborative Filtering is measuring the similarity matrix of the users $B^{(user)}$ (user-based CF) or of the items $B^{(item)}$ (item-based CF).\\ The difference among the models lies in the definition of the similarity; our SSCF makes use of the Sapling Similarity. The other metrics we consider are:\\\\
\textbf{Common Neighbors} \cite{liben2007link}:
\begin{equation}
	B^{CN}_{ij} = CO_{ij}
\end{equation}
\textbf{Jaccard} \cite{jaccard1901etude}:
\begin{equation}
	B^{JA}_{ij} = \frac{CO_{ij}}{k_i+k_j-CO_{ij}}
\end{equation}
\textbf{Adamic/Adar} \cite{adamic2003friends}:
\begin{equation}
	B^{AD}_{ij} = \sum_\lambda\frac{M_{i\lambda}M_{j\lambda}}{log(k_\lambda)}
\end{equation}
\textbf{Resource Allocation Index} \cite{zhou2009predicting}:
\begin{equation}
	B^{RA}_{ij} = \sum_\lambda\frac{M_{i\lambda}M_{j\lambda}}{k_\lambda}
\end{equation}
\textbf{Cosine Similarity} \cite{salton1983introduction}:
\begin{equation}
	B^{CS}_{ij} = \frac{CO_{ij}}{\sqrt{k_ik_j}}
\end{equation}
\textbf{Sorensen index} \cite{sorensen1948method}:
\begin{equation}
	B^{SO}_{ij} = \frac{1}{k_i+k_j}CO_{ij}
\end{equation}
\textbf{Hub depressed index} \cite{ravasz2002hierarchical, hidalgo2007product}:
\begin{equation}
	B^{HDI}_{ij} = \frac{1}{max(k_i,k_j)}CO_{ij}
\end{equation}
\textbf{Hub promoted index} \cite{ravasz2002hierarchical}:
\begin{equation}
	B^{HPI}_{ij} = \frac{1}{min(k_i,k_j)}CO_{ij}
\end{equation}
\textbf{Taxonomy Network} \cite{zaccaria2014taxonomy}:
\begin{equation}
	B^{TN}_{ij} = \frac{1}{max(k_i,k_j)}\sum_\lambda\frac{M_{i\lambda}M_{j\lambda}}{k_\lambda}
\end{equation}
\textbf{Probabilistic Spreading} \cite{zhou2007bipartite}:
\begin{equation}
	B^{ProbS}_{ij} = \frac{1}{k_j}\sum_\lambda\frac{M_{i\lambda}M_{j\lambda}}{k_\lambda}
\end{equation}
\textbf{Pearson Correlation Coefficient} \cite{shardanand1995social}:
\begin{equation}
	B^{PCC}_{ij} = \frac{\sum_\lambda(M_{i\lambda}-k_i/N)(M_{j\lambda}- k_j/N)}{\sqrt{\sum_\lambda(M_{i\lambda}-k_i/N)^2}\sqrt{\sum_\lambda(M_{j\lambda}- k_j/N)^2}}
\end{equation}
\subsection{Model-based Collaborative Filtering}
In section \ref{sec:memory-based} we will compare SSCF with other memory-based CF on the three datasets country-export, Amazon-product, and Milan GPS data. In this comparison, we will also include Non-Negative Matrix Factorization (NMF) \cite{fevotte2011algorithms} and LightGCN \cite{he2020lightgcn}. The former has the goal to decompose the original M matrix ($|U|\times|\Gamma|$) in the product of two matrices L ($|U|\times K$) and R ($K\times|\Gamma|$). Each row of L is the embedding of a user $e_i$ and each column of R is the embedding of an item $e_\alpha$. The recommendation score $S_{i\alpha}$ is the scalar product of $e_i$ and $e_\alpha$. In applying NMF to our three datasets we optimized the embedding size K directly on the test data finding the values K = 7 for country-export, K = 201 for Amazon-Product, and K = 13 on Milan GPS data.\\
LightGCN is a GCN model so, like NMF, its goal is to build embedding representations for users and items and measure the recommendation scores through a scalar product of the embeddings. Following  the procedure adopted by the authors of LightGCN, the hyperparameters that we optimized are the L2 regularization term and the number of layers. The adopted values for the L2 regularization term and the number of layers are respectively: 0.001 and 3 for country-export, 0.001 and 4 for Amazon-product, and 0.01 and 3 for the Milan GPS data.\\
Details about the optimization of the hyperparameters are provided in the supplementary material.

\subsection{Performance indicators}
Following the literature \cite{he2020lightgcn,choi2021lt,choi2022perturbation}, the performance measures we use to evaluate the goodness of the models are:
\begin{itemize}
    \item precision@20: the fraction of elements in the top 20 recommendations that are relevant;
    \item recall@20: the fraction of relevant elements that are in the top 20 recommendations;
    \item ndcg@20: The Normalized Discounted Cumulative Gain \cite{jarvelin2002cumulated} computed considering only the top 20 scores in the ranking.
\end{itemize}
Each metric is computed separately for each user and then the average is reported.\\
Some of the papers with which we compare SSCF in section \ref{sec:model-based} do not use precision@20; for this reason in the comparison on the Gowalla, Yelp2018, and Amazon-Book datasets we will adopt only recall@20 and ndcg@20.

\section{Experimental results}
\subsection{Reproduction and comparison of similarity measures}\label{sec:memory-based}
Table \ref{tab:results_small} shows the accuracy in terms of precision@20, recall@20, and ndcg@20 of various memory-based CFs, based on Sapling Similarity (SSCF) and on the other similarities we presented in section \ref{sec:similarities}.\begin{table}[H]
\centering
\begin{tabular}{|l|l|ccc|}

\hline
\textbf{dataset} & \textbf{model}           &  \textbf{precision@20}   & \textbf{recall@20}   & \textbf{ndcg@20}\\ \hline
\multirow{12}{*}{Country-Export}
& Common neighbors  ($\gamma = 0.8$)       & 0.1689 & 0.0362 & 0.1803       \\  
& Jaccard          ($\gamma = 0.8$)        & 0.1840 & 0.0447 & 0.1988       \\ 
& Adamic/adar      ($\gamma = 0.8$)        & 0.1663 & 0.0355 & 0.1768       \\ 
& Resource allocation  ($\gamma = 0.8$)    & 0.1843 & 0.0417 & 0.1936       \\  
& Cosine similarity   ($\gamma = 0.8$)     & 0.1710 & 0.0419 & 0.1792       \\  
& Sorensen            ($\gamma = 0.8$)     & 0.1743 & 0.0435 & 0.1876       \\ 
& Hub depressed index  ($\gamma = 0.9$)    & 0.1609 & 0.0387 & 0.1759       \\ 
& Hub promoted index   ($\gamma = 0.6$)    & 0.1568 & 0.0354 & 0.1720       \\ 
& Taxonomy network    ($\gamma = 0.9$)     & 0.1790 & 0.0413 & 0.1889       \\ 
& Probabilistic spreading ($\gamma = 0.7$) & 0.1722 & 0.0408 & 0.1843       \\
& Pearson Similarity   ($\gamma = 0.6$)    & 0.1869 & 0.0430 & 0.2041       \\ 
& Sapling Similarity (SSCF)   ($\gamma = 0.7$)   & \textbf{0.2118} & \textbf{0.0479} & \textbf{0.2259}       \\ 
& Preferential attachment  & 0.1157 & 0.0303 & 0.1358       \\ 
& NMF                      & 0.1683 & 0.0354 & 0.1805       \\ 
& LightGCN                 & 0.1766 & 0.0400 & 0.1929       \\ 
\hline
\hline
\multirow{12}{*}{Amazon-Product} 
& Common neighbors    ($\gamma = 0.9$)     & 0.0051 & 0.1019 & 0.0445       \\  
& Jaccard            ($\gamma = 0.9$)      & 0.0058 & 0.1162 & 0.0518      \\ 
& Adamic/adar         ($\gamma = 0.9$)     & 0.0054 & 0.1088 & 0.0478     \\ 
& Resource allocation ($\gamma = 0.8$)     & 0.0058 & 0.1153 & 0.0507      \\  
& Cosine similarity  ($\gamma = 0.9$)      & 0.0055 & 0.1093 & 0.0476      \\  
& Sorensen          ($\gamma = 0.9$)       & 0.0056 & 0.1129 & 0.0493      \\ 
& Hub depressed index   ($\gamma = 0.9$)   & 0.0057 & 0.1139 & 0.0504      \\ 
& Hub promoted index   ($\gamma = 0.8$)    & 0.0046 & 0.0915 & 0.0382      \\ 
& Taxonomy network   ($\gamma = 0.8$)      & 0.0061 & 0.1211 & 0.0542      \\ 
& Probabilistic spreading ($\gamma = 0.6$) & 0.0058 & 0.1155 & 0.0511      \\ 
& Pearson Similarity   ($\gamma = 0.9$)    & 0.0059 & 0.1180 & 0.0528      \\
& Sapling Similarity (SSCF)   ($\gamma = 0.8$)   & \textbf{0.0065} & \textbf{0.1299} & \textbf{0.0593}      \\
& Preferential attachment  & 0.0021 & 0.0410 & 0.0157       \\ 
& NMF                      & 0.0044 & 0.0876 & 0.0392       \\ 
& LightGCN                 & 0.0057 & 0.1132 & 0.0472       \\ 
\hline
\hline
\multirow{12}{*}{Milan GPS data}
& Common neighbors    ($\gamma = 0.3$)     & 0.0217 & 0.0951 & 0.0624      \\  
& Jaccard             ($\gamma = 0.4$)     & 0.0223 & 0.0987 & 0.0650      \\ 
& Adamic/adar        ($\gamma = 0.3$)      & 0.0221 & 0.0972 & 0.0634      \\ 
& Resource allocation  ($\gamma = 0.2$)    & 0.0220 & 0.0953 & 0.0627      \\  
& Cosine similarity    ($\gamma = 0.3$)    & 0.0222 & 0.0967 & 0.0636      \\  
& Sorensen            ($\gamma = 0.4$)     & 0.0223 & 0.0985 & 0.0651      \\ 
& Hub depressed index   ($\gamma = 0.3$)   & 0.0221 & 0.0964 & 0.0641      \\ 
& Hub promoted index    ($\gamma = 0.2$)   & 0.0217 & 0.0939 & 0.0617     \\ 
& Taxonomy network     ($\gamma = 0.3$)    & 0.0220 & 0.0944 & 0.0639      \\ 
& Probabilistic spreading ($\gamma = 0.2$) & 0.0211 & 0.0960 & 0.0640      \\ 
& Pearson Similarity   ($\gamma = 0.3$)    & 0.0217 & 0.0961 & 0.0653     \\
& Sapling Similarity (SSCF)  ($\gamma = 0.4$)    & \textbf{0.0227} & \textbf{0.0995} & \textbf{0.0682}      \\
& Preferential attachment  & 0.0145 & 0.0582 & 0.0403        \\ 
& NMF                      & 0.0174 & 0.0773 & 0.0509       \\ 
& LightGCN                 & 0.0226 & 0.0978 & 0.0670       \\ 
\hline
\end{tabular}
\caption{Comparison among different similarity metrics. The results are achieved by using Collaborative Filtering on the country-export, Amazon-product, and Milan GPS datasets. We indicate the memory-based CFs with the name of the respective similarity metrics we used (our SSCF method is the Sapling Similarity) and the optimized $\gamma$ value. We also show the performance of preferential attachment, NMF, and LightGCN. For each dataset the highest values of each indicator (in bold) are reached by Sapling Similarity.}
\label{tab:results_small}
\end{table}We consider three relatively small datasets: country-export, Amazon-product, and Milan GPS data. The results we present refer to the weighted average between the item-based and the user-based approaches defined in equation \ref{eq:hybrid} (the performance of user-based and item-based models are shown in the supplementary material together with plots about the $\gamma$ optimization for all the similarity metrics). We also show the performance of preferential attachment (a simple model where the probability of having a link between a user and an item is the product of their degree), Non-negative matrix factorization (NMF), and LightGCN.\\
Our results indicate that in these datasets, Sapling Similarity (SSCF) outperforms not only other similarity metrics but also NMF and LightGCN. This means that with this memory-based approach, we can achieve better results than more complex model-based architectures, and this is a significant result also because of the higher simplicity and interpretability of memory-based CF with respect to model-based ones.\\
Before showing that SSCF is competitive also with the state-of-the-art model-based approaches, and considering also benchmark datasets for CFs, we provide a practical example taken from the country-export data. This example shows intuitively how Sapling Similarity is able to capture relevant information with respect to other similarity measures.

\subsection{Zambia, Saudi Arabia, and Japan: a case study}
\begin{figure*}[ht!]
	\centering
		\includegraphics[width=1\textwidth]{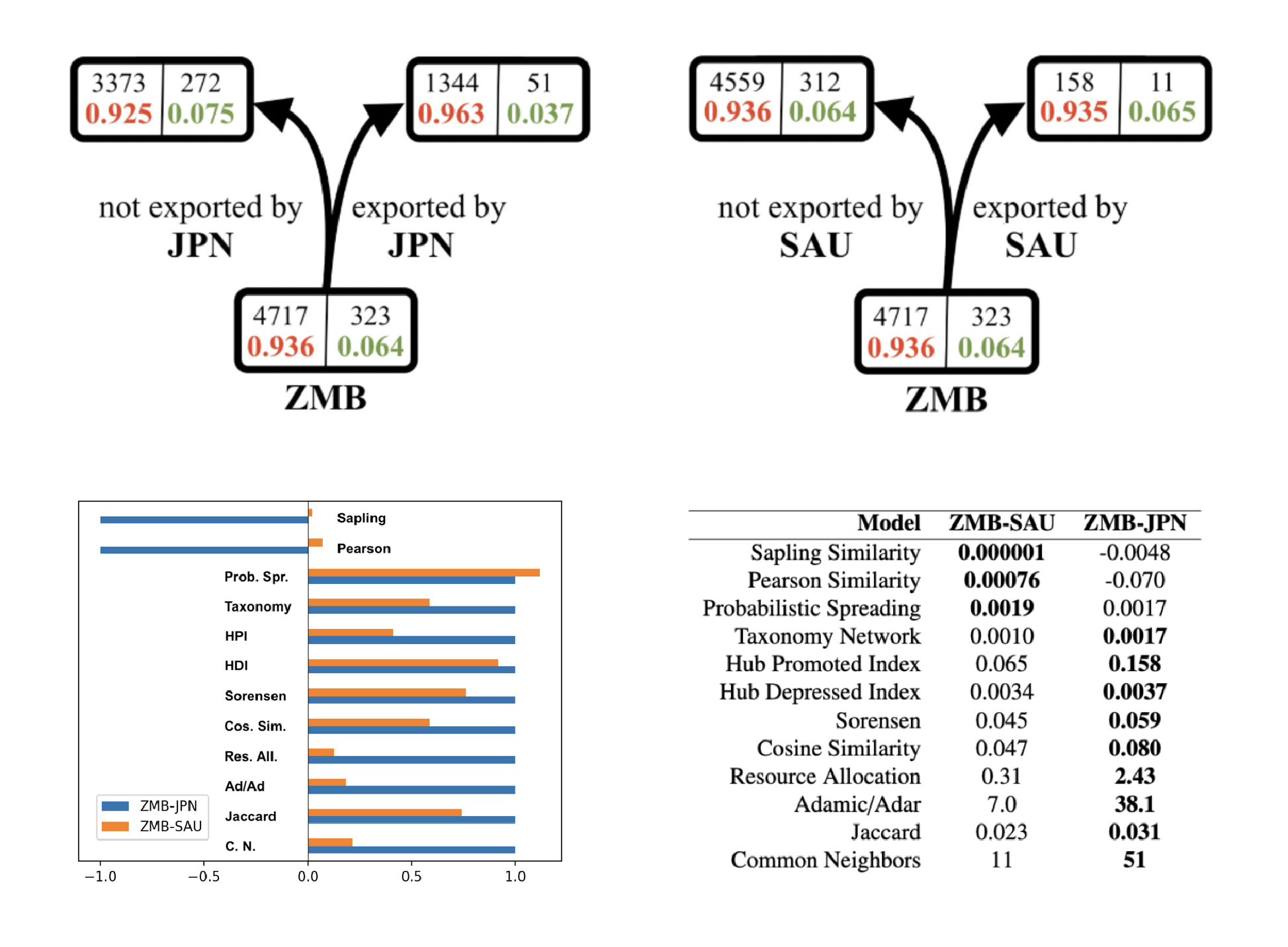}
		\caption{At the top, the Decision Saplings are used to compute the similarity between Zambia and Japan (on the left) and between Zambia and Saudi Arabia (on the right). Knowing that a product is exported by Japan reduces the probability that Zambia exports it, leading to a negative Sapling Similarity; on the contrary, Saudi Arabia does not add any information and so the Similarity is zero. On the bottom right we compare the similarity metrics between Zambia and Japan and between Zambia and Saudi Arabia. On the bottom on the left, the blue bars correspond to the similarity between Zambia and Japan, and the orange ones to the similarity between Zambia and Saudi Arabia. Both values are normalized with respect to the blue bars. Only Sapling and Pearson Similarity correctly quantify the anti-correlation between Zambia and Japan exports.}
		\label{fig:ZMB_SAU_JPN}
\end{figure*}

Here we provide an example of some characteristics of the data which are better captured by the Sapling Similarity by using country-export data. In figure \ref{fig:ZMB_SAU_JPN} we show the Decision Sapling used to compute the similarity between Zambia and Saudi Arabia on the top right and Zambia and Japan on the top left.
If we pick a random product and we ask if Zambia exports it, the probability is 6.4\%. If we ask if Zambia exports a random product chosen from those exported by Japan, the probability drops to 3.7\%. Knowing that Japan exports something reduces our confidence in the fact that also Zambia can export it since the two countries are anti-correlated. To express this concept we need to define a negative similarity between Zambia and Japan so that in equation \ref{eq:density_user}, Japan gives a negative contribution to the recommendation of a product to Zambia. The economic reason is that Japan is specialized in technological complex products, while Zambia is more focused on raw materials and does not have the capability to export the same products as Japan. If we apply the same argument to the case of Saudi Arabia and Zambia, whose Decision Sapling is represented in the top right, knowing that the former exports something does not add information about a possible export of the latter. In this sense, Zambia and Saudi Arabia are uncorrelated and their similarity should be close to zero, so that in equation \ref{eq:density_user}, Saudi Arabia does not give any information on a possible export of Zambia. For the sake of completeness, the most similar countries to Zambia result to be African countries like Tanzania, Zimbabwe, and Uganda. \\
With an approach based only on co-occurrences the difference between Japan, Saudi Arabia, and Zambia is practically negligible. In the bottom part of figure \ref{fig:ZMB_SAU_JPN} we show how the similarity between Japan and Zambia is either higher or comparable to the one between Saudi Arabia and Zambia when we use classic similarity metrics. The only metrics that are able to understand the real difference between Zambia and Japan are Sapling and Pearson Similarity because of the allowance of negative values.\\
\subsection{Comparison of CFs on benchmark datasets} \label{sec:model-based}
In this section, we present the results of a comparison between the state-of-the-art, model-based CFs and our SSCF. We use the three datasets Gowalla, Yelp2018, and Amazon-Book, which are widely used to compare CF methods.
\begin{table*}[ht!]
\centering
\begin{tabular}{|r|cc|cc|cc|}
\hline
   &\multicolumn{2}{c|}{\textit{Gowalla}}&\multicolumn{2}{c|}{\textit{Yelp2018}}&\multicolumn{2}{c|}{\textit{Amazon-Book}}\\
\hline
\textbf{model}              & \textbf{recall@20}   & \textbf{ndcg@20}   & \textbf{recall@20}   & \textbf{ndcg@20}&   \textbf{recall@20}   & \textbf{ndcg@20}\\ \hline
NGCF               & 0.1570 & 0.1327 & 0.0579 & 0.0477 & 0.0344 & 0.0263         \\
LightGCN           & 0.1830 & 0.1554 & 0.0649 & 0.0530 & 0.0411 & 0.0315         \\
UltraGCN           & 0.1862 & 0.1580 & 0.0683 & 0.0561 & 0.0681 & 0.0556         \\
simpleX            & 0.1872 & 0.1557 & 0.0701 & 0.0575 & 0.0583 & 0.0468         \\
LT-OCF             & 0.1875 & 0.1574 & 0.0671 & 0.0549 & 0.0442 & 0.0341         \\
BSPM-LM            & 0.1901 & 0.1548 & 0.0713 & 0.0584 & 0.0733 & 0.0609         \\
BSPM-EM            & \textbf{0.1920} & \textbf{0.1597} & \textbf{0.0720} & \textbf{0.0593} & 0.0733 & 0.0610         \\
Sapling UB         & 0.1640 & 0.1346 & 0.0556 & 0.0459 & 0.0504 & 0.0406         \\
Sapling IB         & 0.1260 & 0.0900 & 0.0555 & 0.0446 & 0.0766 & \textbf{0.0648}\\
SSCF                & 0.1775 & 0.1390 & 0.0664 & 0.0542 & \textbf{0.0773} & 0.0647\\
\hline
\end{tabular}
\caption{Accuracy in terms of recall@20 and ndcg@20 on Gowalla, Yelp2018, and Amazon-Book dataset achieved by SSCF and current state-of-the-art models. For each dataset, we bolded the highest values of each indicator. SSCF achieved excellent results considering that it is a highly interpretable method. Moreover, in the case of the Amazon-Book dataset, it is even the new state-of-the-art model. We also show in the table the performance of the user-based and the item-based Sapling Similarity.}
\label{tab:results_big}
\end{table*}
In Table \ref{tab:results_big}, we show the values for recall@20 and ndcg@20. The optimization of the $\gamma$ parameter is shown in the supplementary material; we find the values 0.5 for Gowalla, 0.8 for Yelp2018, and 0.8 for Amazon-Book. In all datasets, SSCF provides better recommendations than NGCF. In the case of Yelp2018, its performance even surpasses LightGCN. However, the most notable result is in the Amazon-Book dataset (the largest one): here SSCF outperforms all existing models and represents the new state-of-the-art.\\
We would like to emphasize that the performance of SSCF does not depend on any hyperparameter that has been optimized on test data, the only parameter $\gamma$ is estimated from train data. In contrast, all the model-based approaches shown here depend on hyperparameters that have been directly optimized on the test set. In a real-world scenario where the ground truth is unknown, the scores of these model-based methods may be lower. To give an example, on the Amazon-Book dataset, we observe that the item-based Sapling Similarity model reaches an ndcg@20 value that is a little higher than the one of the SSCF, even if they are very close. Moreover, if we would have optimized the $\gamma$ parameter on the test set we would have found 0.9 instead of 0.8. In this case, SSCF would have reached a recall@20 of 0.0779 and an ndcg@20 of 0.0654. This example shows that if we use test data to optimize hyperparameters, the results we show may be overestimated compared to a real scenario.

\section{Conclusions and future works}
This paper introduces the SSCF, a memory-based collaborative filtering based on a novel metric of similarity between nodes (users or items) in weightless bipartite networks, the Sapling Similarity. This metric is built using concepts from information theory, with a probabilistic approach. The main novelty introduced by the Sapling Similarity is the permission of negative values, that are not conceived in the case of classical similarity metrics based on co-occurrences. Moreover, our probabilistic approach naturally requires (and introduces in the mathematical formulation) the total size $N$ of the layer (users or items) under investigation. The idea is to look at how the probability that a user is connected to an item (or vice versa) changes if we know that this item is connected to another user. If the probability decreases, then the two users are anti-correlated and we assign a negative similarity; if the probability does not change the similarity of the two users is zero; finally, if the probability increases they are positively correlated and their similarity is positive. This criterion is expressed in mathematical terms by using the variation of Gini Impurity, following the ideas at the base of Decision Trees and Random Forest \cite{breiman1984cart,breiman2001random}.\\
Similarity metrics are widely used to build recommender systems with the aim of suggesting new items to users, in particular in the field of collaborative filtering. Here we show that memory-based collaborative filtering based on Sapling Similarity outperforms the ones based on other similarity metrics.\\
With the always-increasing interest in machine learning, many model-based collaborative filterings have been introduced in the last decade. In this study, we define the SSCF method as a weighted average of a user-based and an item-based approach based on Sapling Similarity. Using three widely used datasets to evaluate collaborative filterings we show that SSCF is competitive with the current state-of-the-art model-based approaches. Moreover in the Amazon-book dataset, our SSCF achieves higher scores than all existing models. This represents a remarkable result given the high interpretability and simplicity of SSCF, whose performance rely on a single easily interpretable hyperparameter, in contrast to model-based approaches such as graph convolution networks, whose results are less interpretable and depend on multiple hyperparameters, often optimized directly by maximizing the model's performance on test data.\\
It is important to note that Sapling Similarity can be used for more than just constructing collaborative filtering. Measuring the similarity between nodes in a bipartite network is equivalent to doing a projection on one of the two layers. Bipartite network projections have many applications, for instance, they can be used to do community detection and clustering in order to reveal the hidden relations among the nodes in the network \cite{saracco2017inferring,tumminello2012identification}. For instance, in \cite{bass2013using} Bass et al use similarity metrics to investigate biological networks.\\
Similarity measures and item-based collaborative filtering are widely used in Economic Complexity to measure the relatedness between countries and products \cite{hidalgo2018principle}, however recent studies \cite{tacchella2021relatedness, albora2022machine} showed that tree-based machine learning algorithms like Random Forest \cite{breiman2001random} and XGBoost \cite{friedman2001greedy,chen2016xgboost} provide better results. The disadvantages of using machine learning with respect to item-based collaborative filtering are the loss of interpretability of the results and the high computational time required to train an algorithm (a training sample is a country with 5040 binary features, one for each product). A possible development in this direction is to use the Sapling Similarity to perform a feature selection in order to reduce the features of the training samples. This would reduce the computational time required to train the algorithms and would also increase the interpretability of the models. Note that positive-definite similarity metrics can not be used for this task, since for a machine learning algorithm the negative relations between products are key \cite{albora2022machine}.\\
The code for our proposed SSCF is available at https://github.com/giamba95/SaplingSimilarity.

\bibliographystyle{vancouver}
\bibliography{sample}
\newpage
\begin{center}
\section*{\fontsize{15}{15}\selectfont Supplementary Information}\vspace{0.5cm}
\end{center}

\section*{S1. Country-export, Amazon-product, and Milan GPS datasets}
Country-export data are derived from the UN-COMTRADE (comtrade.un.org) dataset consisting of export data about 169 countries and 5040 products. A common procedure in the economic complexity literature \cite{tacchella2012new,albora2023product} is to compute the Revealed Comparative Advantage (RCA) introduced by Balassa \cite{balassa1965trade} defined as:
\begin{equation}
	RCA_{i\alpha}=\frac{E_{i\alpha}}{\sum_{\lambda}E_{i\lambda}}\frac{\sum_{l}\sum_{\lambda}E_{l\lambda}}{\sum_{l}E_{l\alpha}}
\end{equation}
where $E_{i\alpha}$ is the volume of product $\alpha$ expressed in american dollars that country $i$ exports.\\
Since RCA is an estimate of how much a country is competitive in the export of a product, with a threshold on the RCA values we can build a binary matrix $M$ representing the bi-adjacency matrix of the bipartite network where a link $(i,\alpha)$ means that country $i$ is competitive in the export of product $\alpha$.
\begin{equation}
	M_{i\alpha} =
	\begin{cases}
		1 ~~~\text{ if } RCA_{i\alpha}\geq1\\[3mm]
		0 ~~~\text{ if } RCA_{i\alpha}<1
	\end{cases}	
\end{equation}
The Amazon-product dataset (http://jmcauley.ucsd.edu/data/amazon) contains the rating data of users for multiple Amazon products. The ratings range from 1 to 5, and they are collected in a matrix $R$. To build the matrix $M$ we select the links with scores equal to or above 3.
\begin{equation}
	M_{i\alpha} =
	\begin{cases}
		1 ~~~\text{ if } R_{i\alpha}\geq3\\[3mm]
		0 ~~~\text{ if } R_{i\alpha}<3
	\end{cases}	
\end{equation}
We removed from the dataset the users and the products with degree less than 10.\\
Finally, the Milan GPS data connects inhabitants with Points Of Interest (POI) in Milan extracted from OpenStreetMap (https://www.openstreetmap.org/). The location data are provided by Cuebiq Inc. (https://www.cuebiq.com/about/data-for-good/). We defined hexagonal areas \cite{biazzo2019general} in the city so that there is a link between an inhabitant and an area if the inhabitant visited a POI located in the area.

\section*{S2. User-based and item-based results on country-export, Amazon-product, and Milan GPS data}
In table \ref{tab:results_item_user} we show the accuracy in terms of precision@20, recall@20, and ndcg@20 of item-based and user-based collaborative filtering based on different similarities. For each indicator, the collaborative filtering based on Sapling Similarity reaches the highest scores. Looking at the table one can norice that, in the item-based case, the Sapling Similarity improvement with respect to the other models is more evident. In the country-export dataset, in the user-based case, Sapling Similarity improves the recall@20 of the second best model by 3.7\%, while in the item-based case, the improvement is 9.5\%. In the Amazon-product dataset, the improvement is 1.7\% in the user-based case and 10.4\% in the item-based one. In the Milan GPS data, the improvement is 0.2\% in the user-based case and 11.6\% in the item-based one.\\
Finally, we also notice that in the country-export and Milan GPS data the user-based approach works better than the item-based one, but in the Amazon-product dataset this depends on the similarity: for instance, Sapling Similarity works better in the item-based case, while COsine Similarity works better in the user-based one.
\begin{table*}[h!]
\centering
\resizebox{\textwidth}{!}{%
\begin{tabular}{|l|l|ccc|ccc|}
\hline
& &\multicolumn{3}{c|}{\textit{User-Based}}&\multicolumn{3}{c|}{\textit{Item-Based}}\\
\hline
& \textbf{model}           &  \textbf{precision@20}   & \textbf{recall@20}   & \textbf{ndcg@20}  &  \textbf{precision@20}   & \textbf{recall@20}   & \textbf{ndcg@20} \\ \hline
\multirow{12}{*}{Country-Export}
& Common neighbors         & 0.1429 & 0.0319 & 0.1635 & 0.0799 & 0.0157 & 0.0743        \\  
& Jaccard                  & 0.1621 & 0.0417 & 0.1788 & 0.1426 & 0.0291 & 0.1380        \\ 
& Adamic/adar              & 0.1414 & 0.0312 & 0.1629 & 0.0852 & 0.0173 & 0.0791        \\ 
& Resource allocation      & 0.1370 & 0.0302 & 0.1562 & 0.0947 & 0.0202 & 0.0892        \\  
& Cosine similarity        & 0.1562 & 0.0400 & 0.1739 & 0.0811 & 0.0156 & 0.0746        \\  
& Sorensen                 & 0.1601 & 0.0416 & 0.1762 & 0.1166 & 0.0227 & 0.1111        \\ 
& Hub depressed index      & 0.1580 & 0.0406 & 0.1745 & 0.1358 & 0.0274 & 0.1313        \\ 
& Hub promoted index       & 0.1432 & 0.0331 & 0.1611 & 0.0473 & 0.0100 & 0.0458        \\ 
& Taxonomy network         & 0.1562 & 0.0409 & 0.1725 & 0.1592 & 0.0348 & 0.1601        \\ 
& Probabilistic spreading  & 0.1479 & 0.0399 & 0.1622 & 0.0947 & 0.0202 & 0.0892        \\
& Pearson Similarity       & 0.1846 & 0.0435 & 0.2018 & 0.1654 & 0.0324 & 0.1803        \\ 
& Sapling Similarity       & \textbf{0.1941} & \textbf{0.0451} & \textbf{0.2107} & \textbf{0.1888} & \textbf{0.0381} & \textbf{0.2064}        \\ 
\hline
\hline
\multirow{12}{*}{Amazon-Product} 
& Common neighbors         & 0.0040 & 0.0794 & 0.0328 & 0.0031 & 0.0629 & 0.0246        \\  
& Jaccard                  & 0.0045 & 0.0894 & 0.0374 & 0.0051 & 0.1026 & 0.0444        \\ 
& Adamic/adar              & 0.0041 & 0.0825 & 0.0338 & 0.0036 & 0.0716 & 0.0295        \\ 
& Resource allocation      & 0.0043 & 0.0866 & 0.0355 & 0.0039 & 0.0789 & 0.0326        \\  
& Cosine similarity        & 0.0043 & 0.0856 & 0.0357 & 0.0035 & 0.0699 & 0.0290        \\  
& Sorensen                 & 0.0044 & 0.0887 & 0.0370 & 0.0047 & 0.0931 & 0.0395        \\ 
& Hub depressed index      & 0.0045 & 0.0902 & 0.0378 & 0.0053 & 0.1057 & 0.0457        \\ 
& Hub promoted index       & 0.0039 & 0.0787 & 0.0331 & 0.0023 & 0.0461 & 0.0177        \\ 
& Taxonomy network         & 0.0050 & 0.1008 & 0.0428 & 0.0055 & 0.1095 & 0.0484        \\ 
& Probabilistic spreading  & 0.0051 & 0.1016 & 0.0434 & 0.0039 & 0.0789 & 0.0326        \\ 
& Pearson Similarity       & 0.0052 & 0.1049 & 0.0444 & 0.0051 & 0.1029 & 0.0459        \\
& Sapling Similarity       & \textbf{0.0053} & \textbf{0.1067} & \textbf{0.0450} & \textbf{0.0060} & \textbf{0.1209} & \textbf{0.0552}        \\
\hline
\hline
\multirow{12}{*}{Milan GPS data}
& Common neighbors         & 0.0208 & 0.0899 & 0.0590 & 0.0084 & 0.0383 & 0.0237       \\  
& Jaccard                  & 0.0213 & 0.0919 & 0.0609 & 0.0118 & 0.0534 & 0.0350       \\ 
& Adamic/adar              & 0.0214 & 0.0927 & 0.0603 & 0.0082 & 0.0382 & 0.0237       \\ 
& Resource allocation      & 0.0212 & 0.0924 & 0.0604 & 0.0081 & 0.0380 & 0.0235       \\  
& Cosine similarity        & 0.0211 & 0.0913 & 0.0604 & 0.0090 & 0.0414 & 0.0261       \\  
& Sorensen                 & 0.0211 & 0.0913 & 0.0605 & 0.0106 & 0.0482 & 0.0310       \\ 
& Hub depressed index      & 0.0211 & 0.0912 & 0.0608 & 0.0128 & 0.0566 & 0.0364       \\ 
& Hub promoted index       & 0.0208 & 0.0892 & 0.0593 & 0.0064 & 0.0306 & 0.0190       \\ 
& Taxonomy network         & 0.0215 & 0.0930 & 0.0618 & 0.0125 & 0.0553 & 0.0355       \\ 
& Probabilistic spreading  & 0.0214 & 0.0923 & 0.0618 & 0.0081 & 0.0380 & 0.0235       \\ 
& Pearson Similarity       & 0.0220 & 0.0937 & 0.0623 & 0.0108 & 0.0497 & 0.0324       \\
& Sapling Similarity       & \textbf{0.0221} & \textbf{0.0939} & \textbf{0.0631} & \textbf{0.0136} & \textbf{0.0617} & \textbf{0.0431}       \\
\hline
\end{tabular}}
\caption{Results achieved with item-based and user-based Collaborative Filtering on the country-export, Amazon-product, and Milan GPS dataset. For each dataset the highest values of each indicator (in bold) are reached by Sapling Similarity.}
\label{tab:results_item_user}
\end{table*}
\section*{S3. Optimization of $\gamma$ in country-export, Amazon-product, and Milan GPS data}
\begin{figure}[H]
	\centering
        \includegraphics[width=0.29\textwidth]{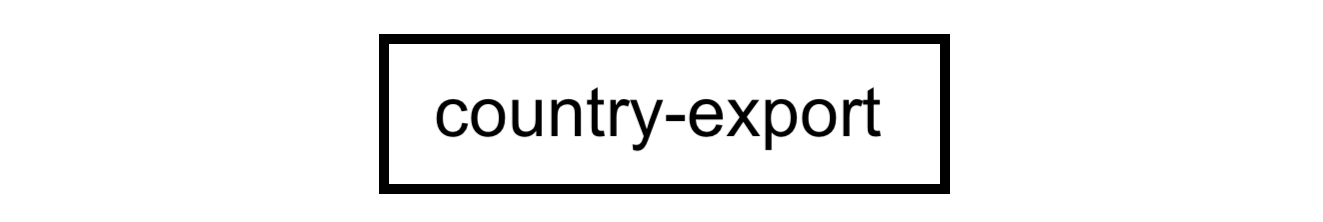}
	\includegraphics[width=0.29\textwidth]{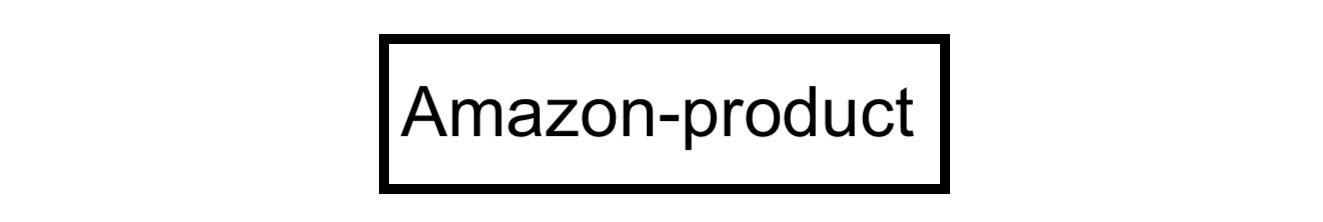}
	\includegraphics[width=0.29\textwidth]{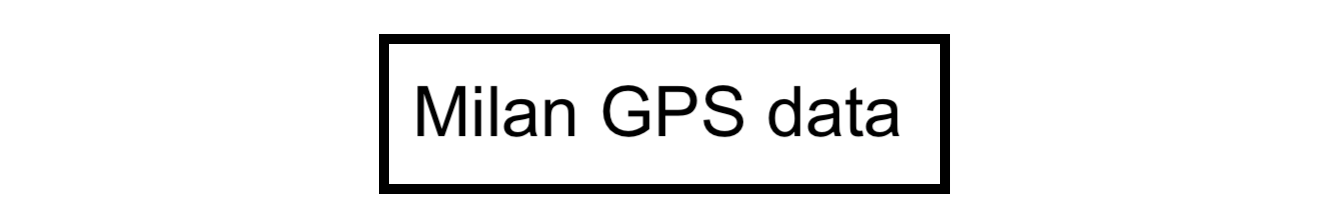}\\
	\includegraphics[width=0.29\textwidth]{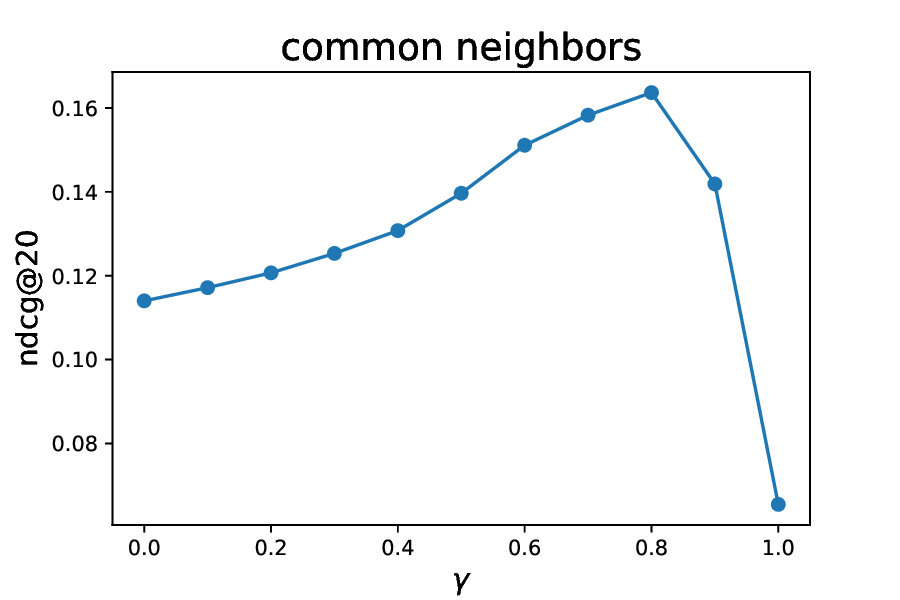}
	\includegraphics[width=0.29\textwidth]{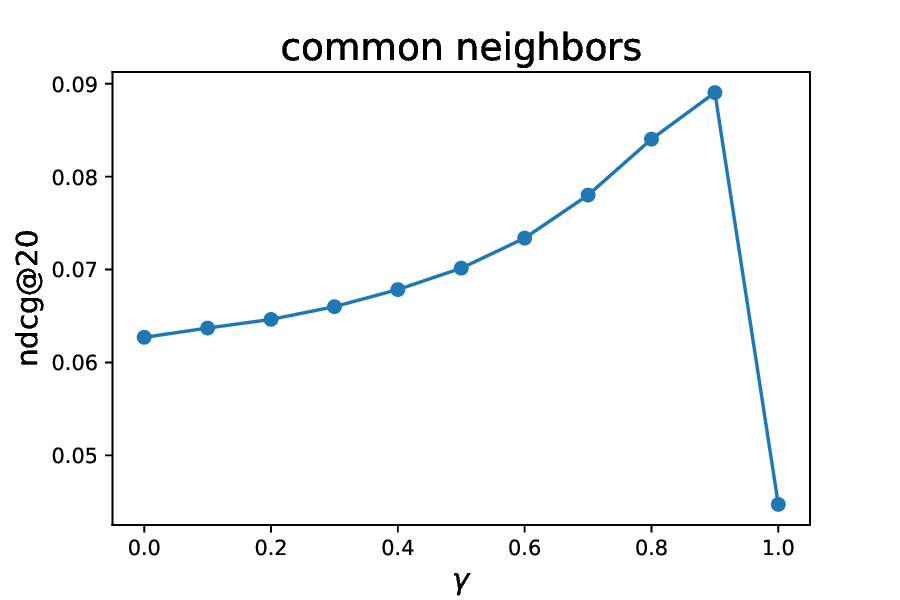}
	\includegraphics[width=0.29\textwidth]{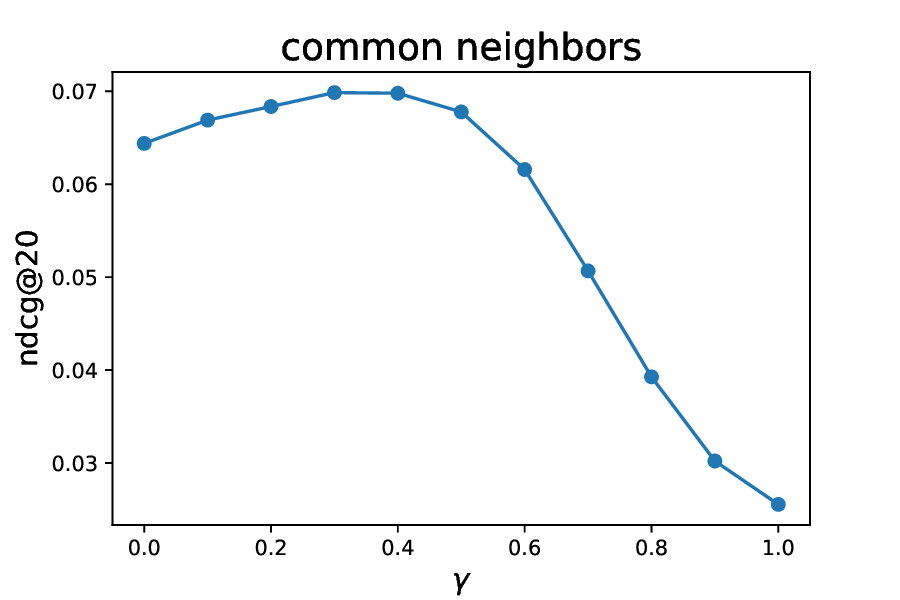}\\
	\includegraphics[width=0.29\textwidth]{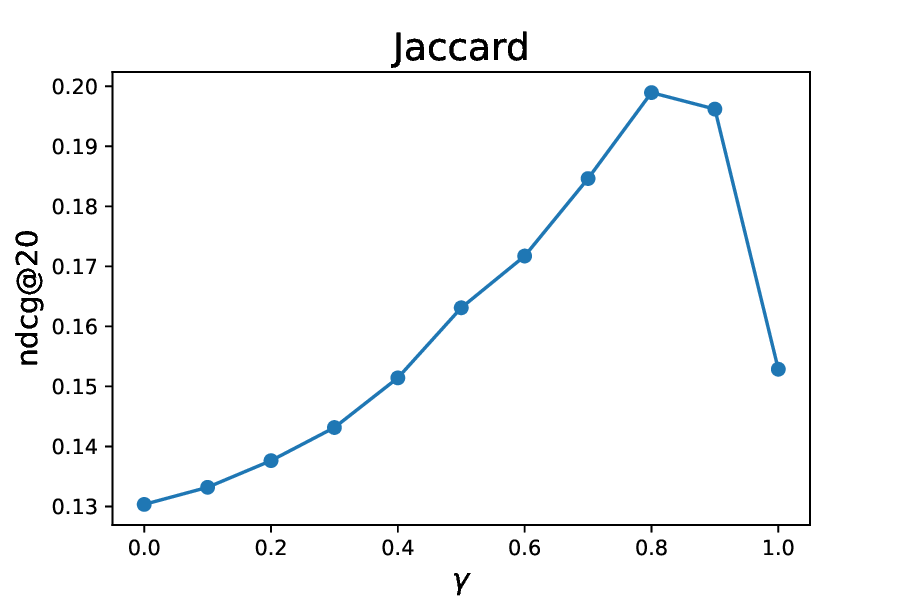}
	\includegraphics[width=0.29\textwidth]{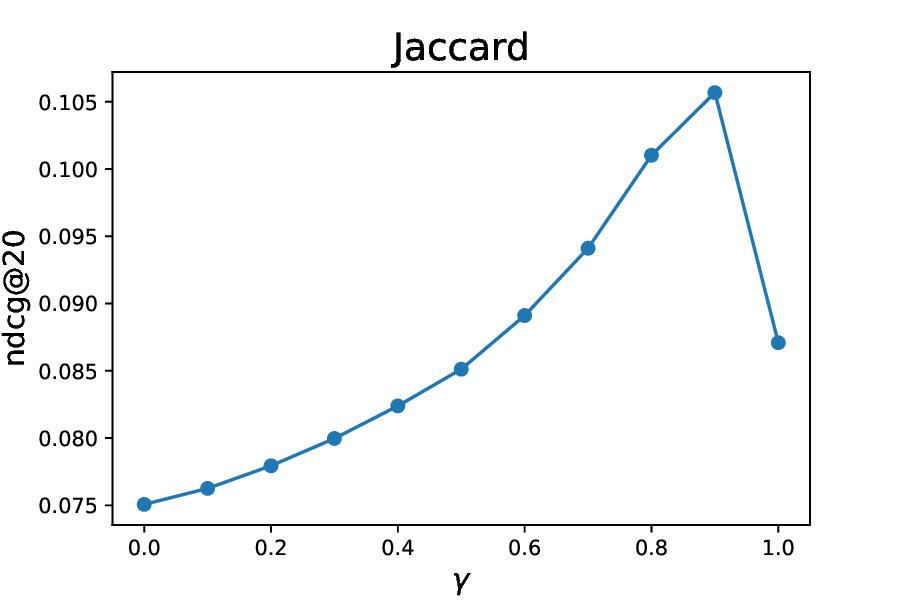}
	\includegraphics[width=0.29\textwidth]{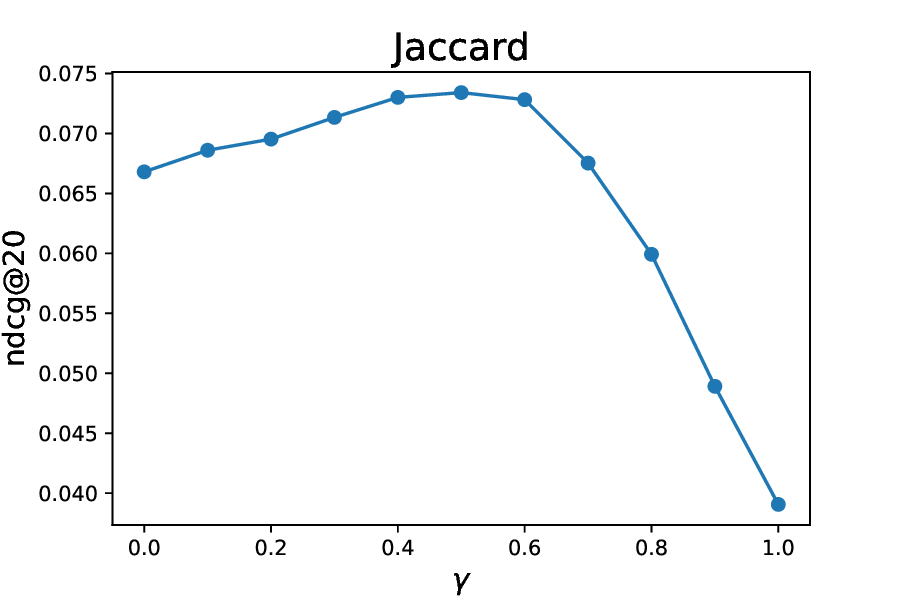}\\
	\includegraphics[width=0.29\textwidth]{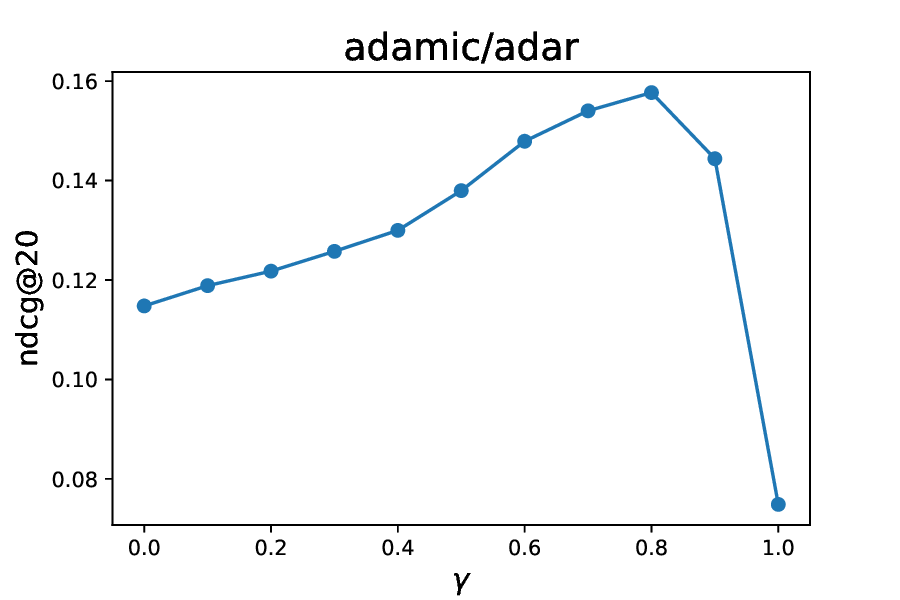}
	\includegraphics[width=0.29\textwidth]{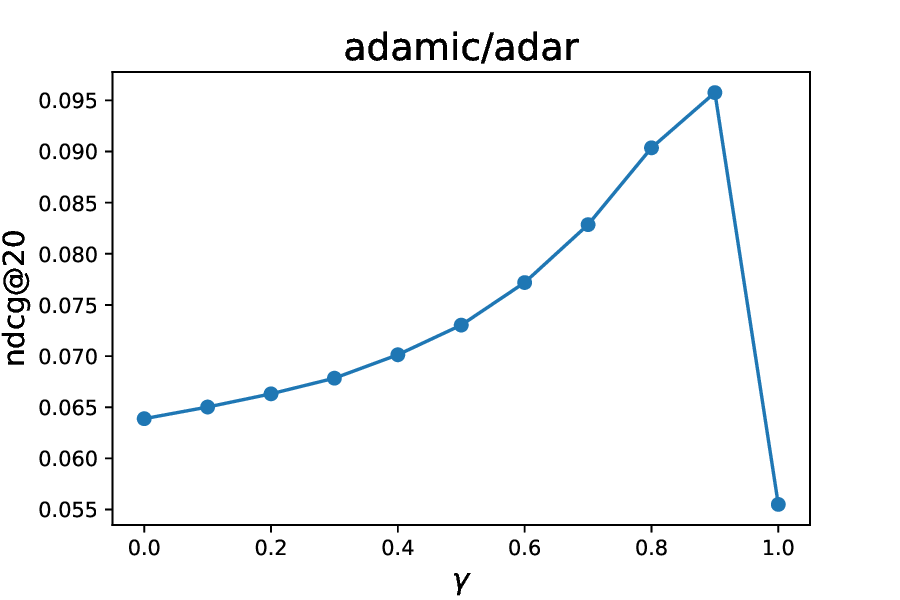}
	\includegraphics[width=0.29\textwidth]{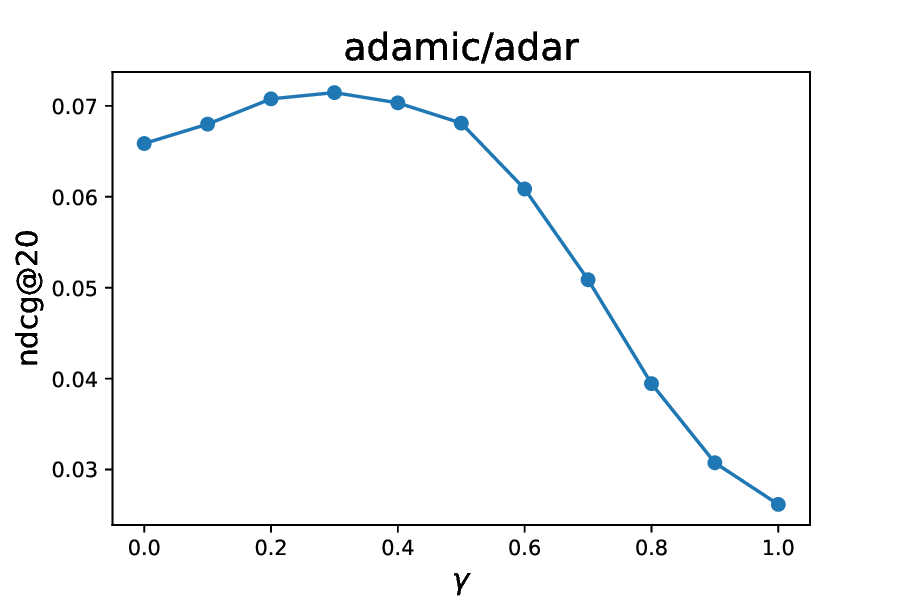}\\
	\includegraphics[width=0.29\textwidth]{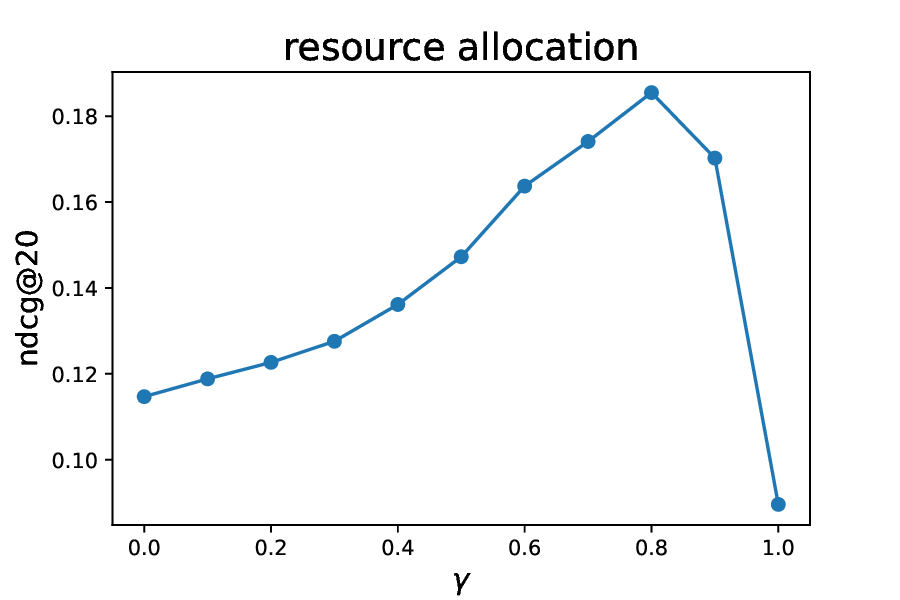}
	\includegraphics[width=0.29\textwidth]{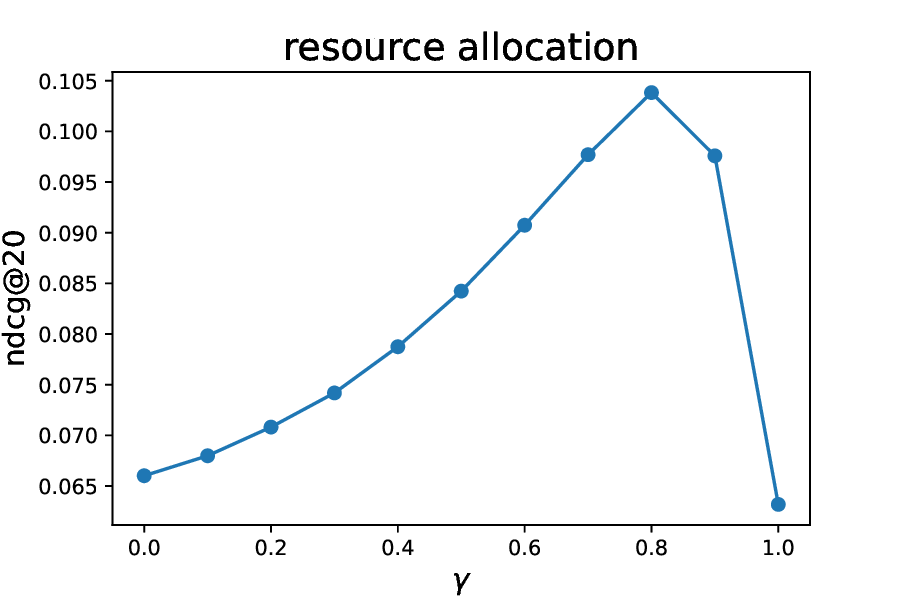}
	\includegraphics[width=0.29\textwidth]{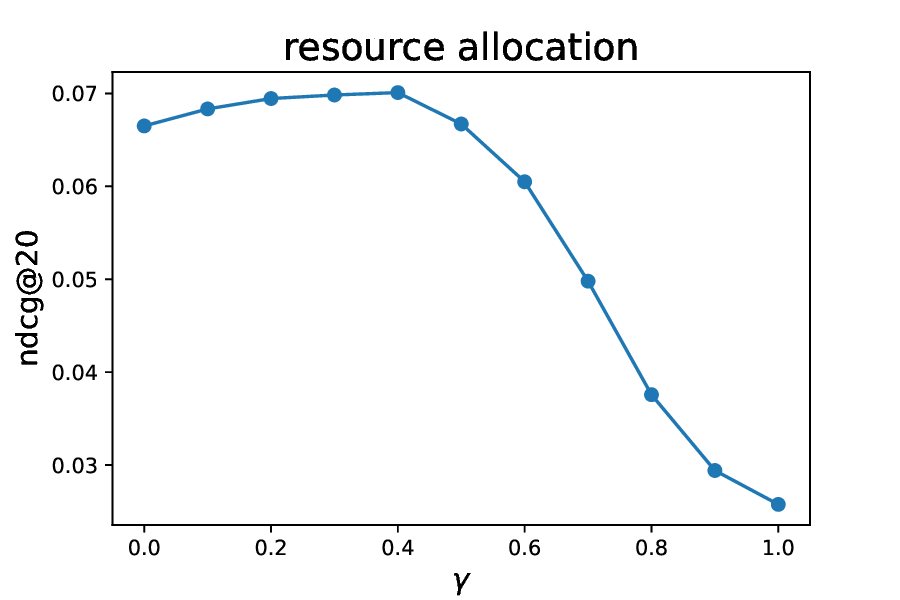}\\
	\includegraphics[width=0.29\textwidth]{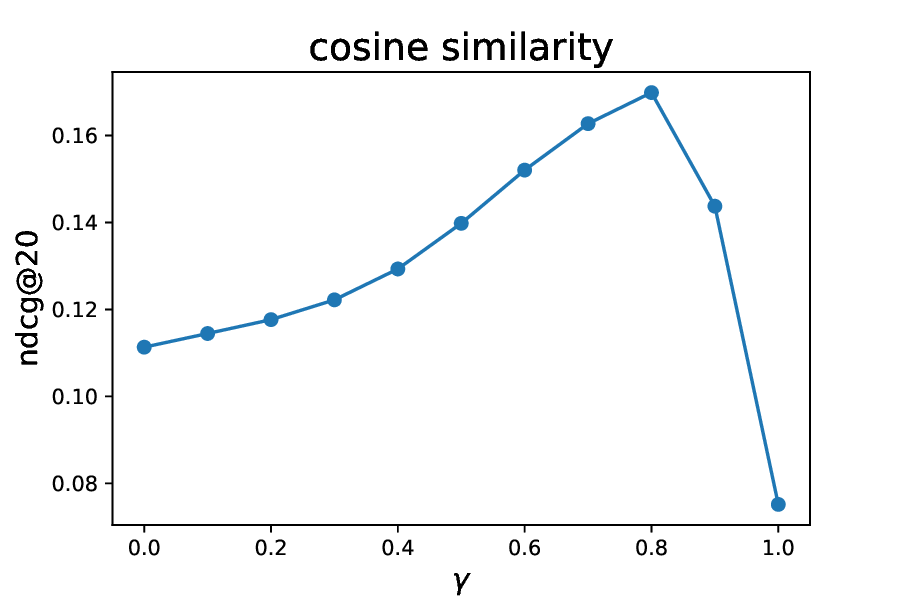}
	\includegraphics[width=0.29\textwidth]{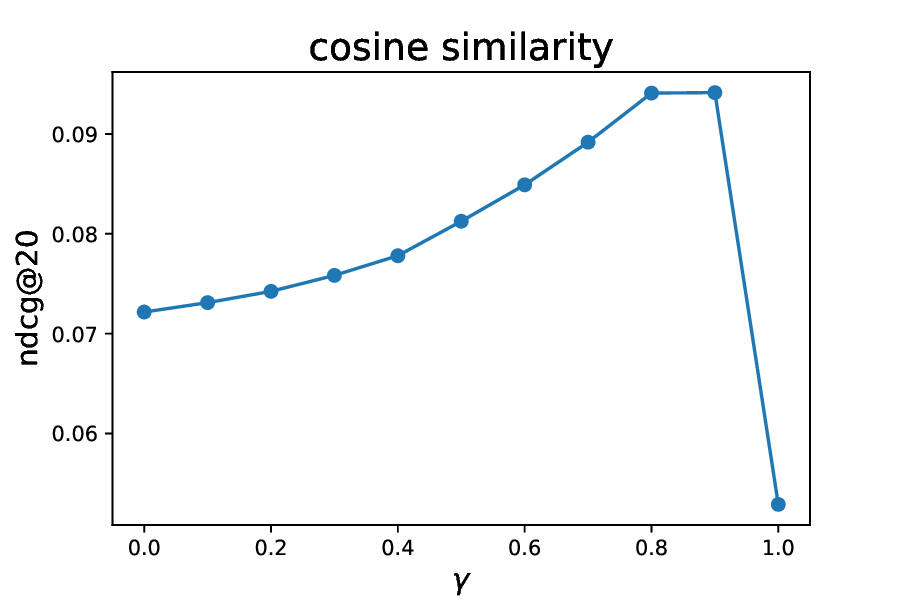}
	\includegraphics[width=0.29\textwidth]{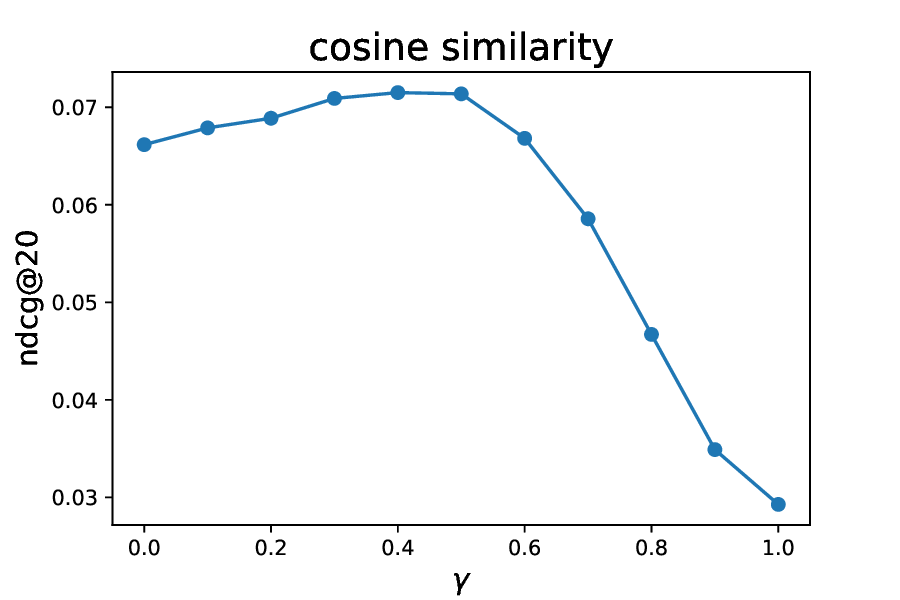}\\
	\caption{Optimization of the $\gamma$ parameter used to combine user-based and item-based approaches in country-export (first column), Amazon-product (second column), and Milan GPS data (third column).}
	\label{fig:gamma_small1}
\end{figure}

\begin{figure}[H]
	\centering
        \includegraphics[width=0.3\textwidth]{Figure/d1.png}
	\includegraphics[width=0.3\textwidth]{Figure/d2.png}
	\includegraphics[width=0.3\textwidth]{Figure/d3.png}\\
        \includegraphics[width=0.3\textwidth]{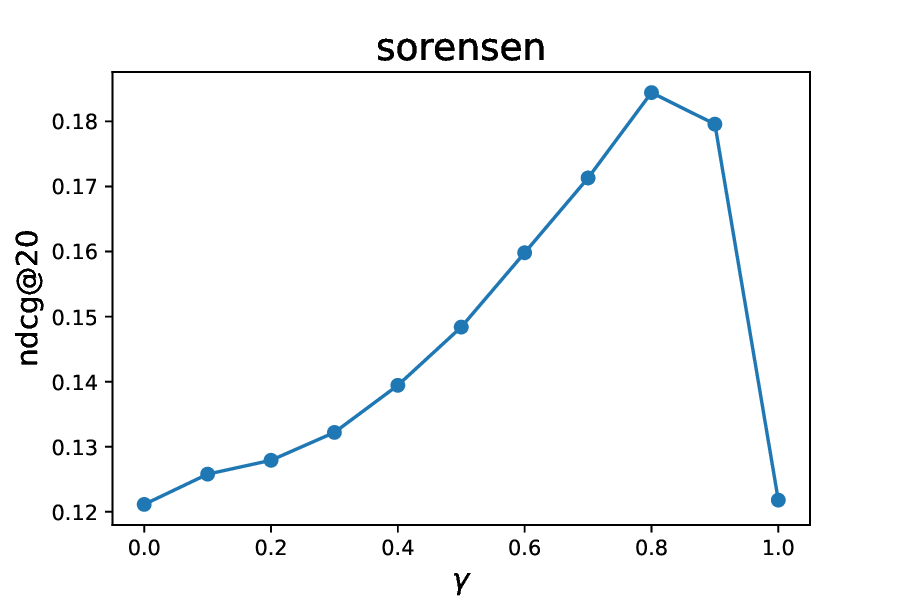}
	\includegraphics[width=0.3\textwidth]{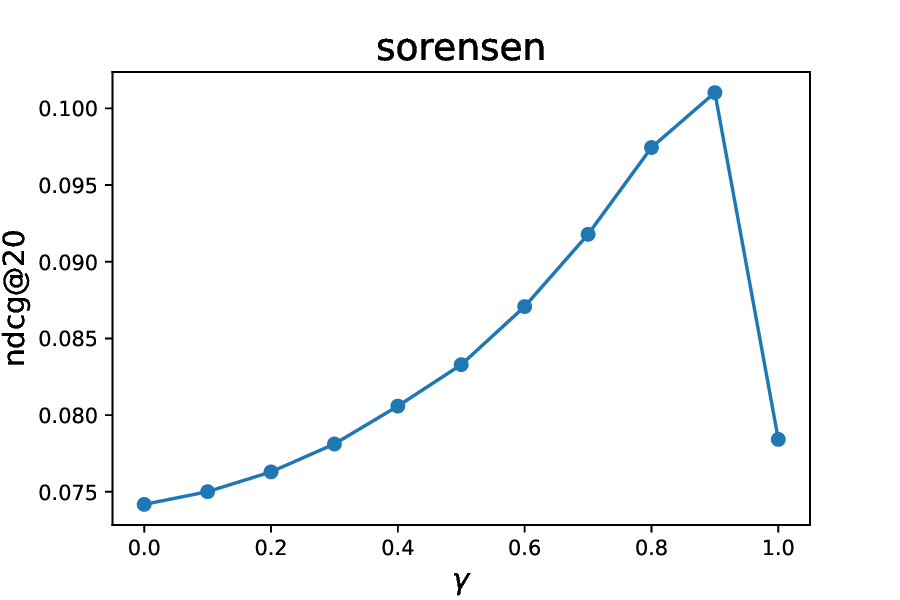}
	\includegraphics[width=0.3\textwidth]{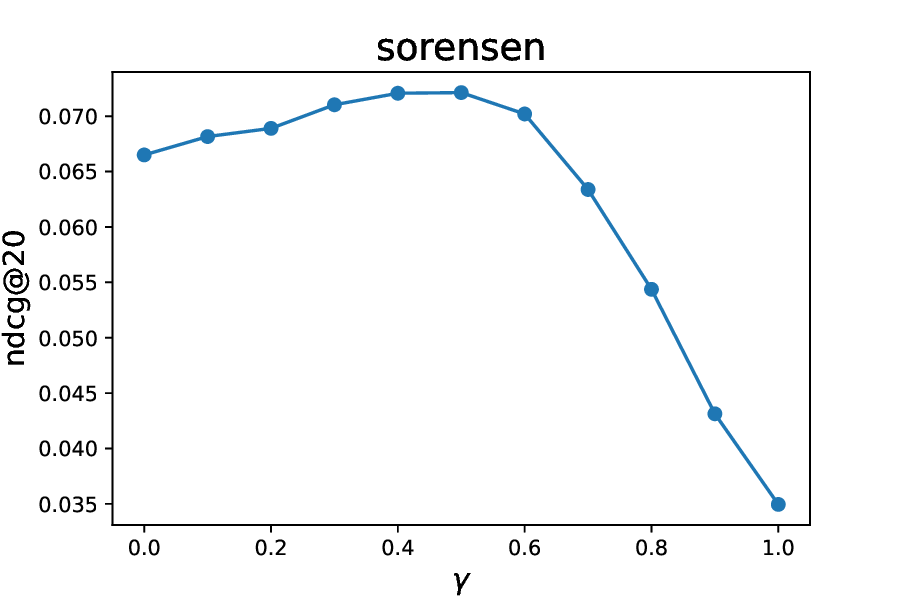}\\
	\includegraphics[width=0.3\textwidth]{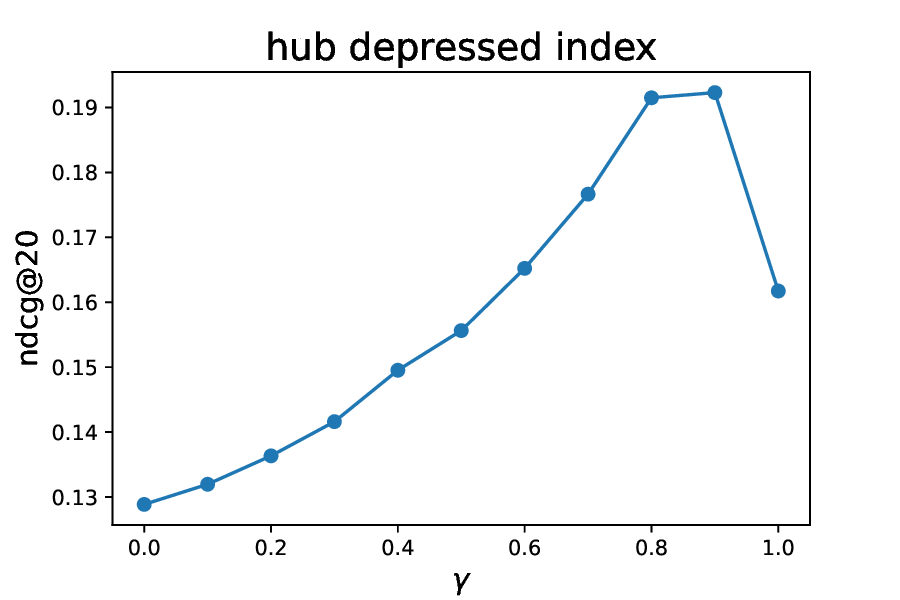}
	\includegraphics[width=0.3\textwidth]{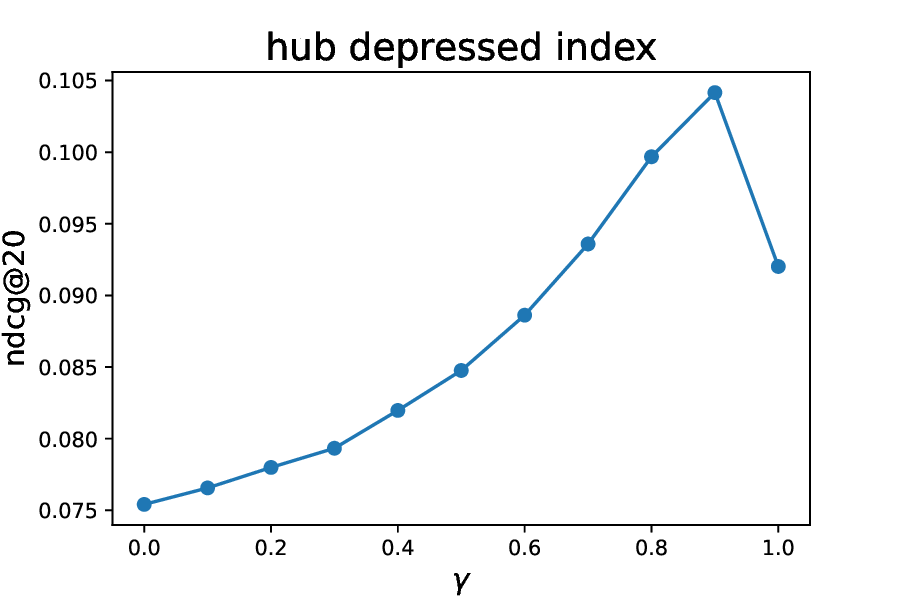}
	\includegraphics[width=0.3\textwidth]{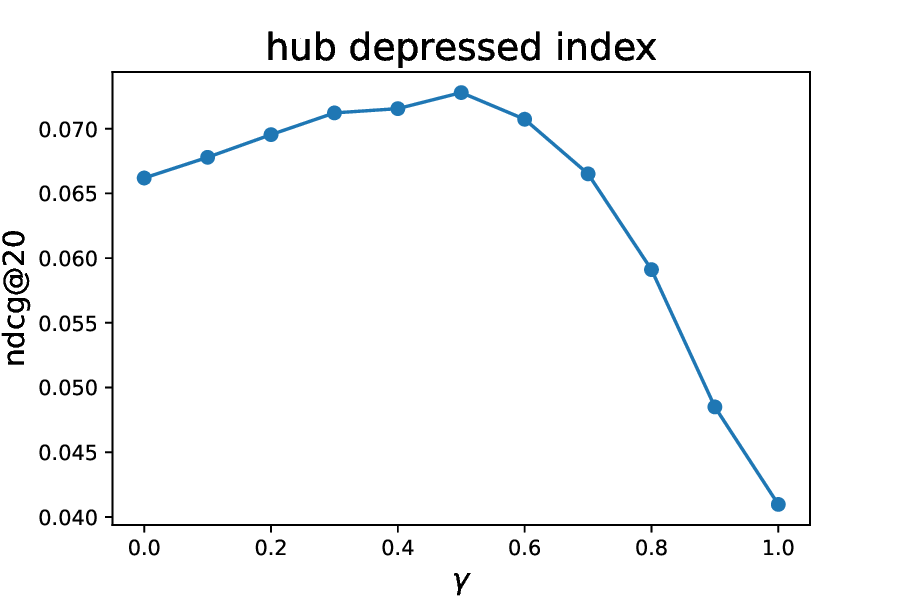}
	\includegraphics[width=0.3\textwidth]{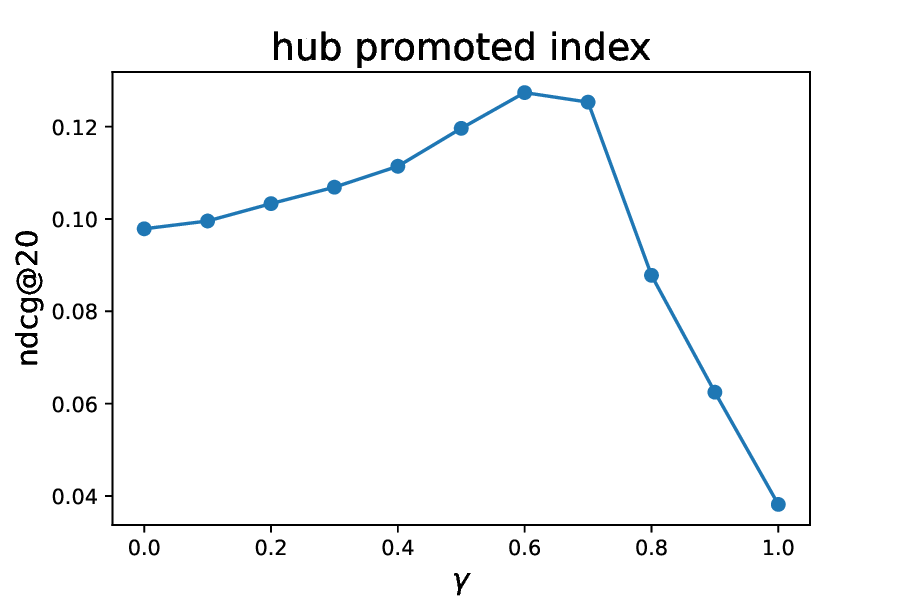}
	\includegraphics[width=0.3\textwidth]{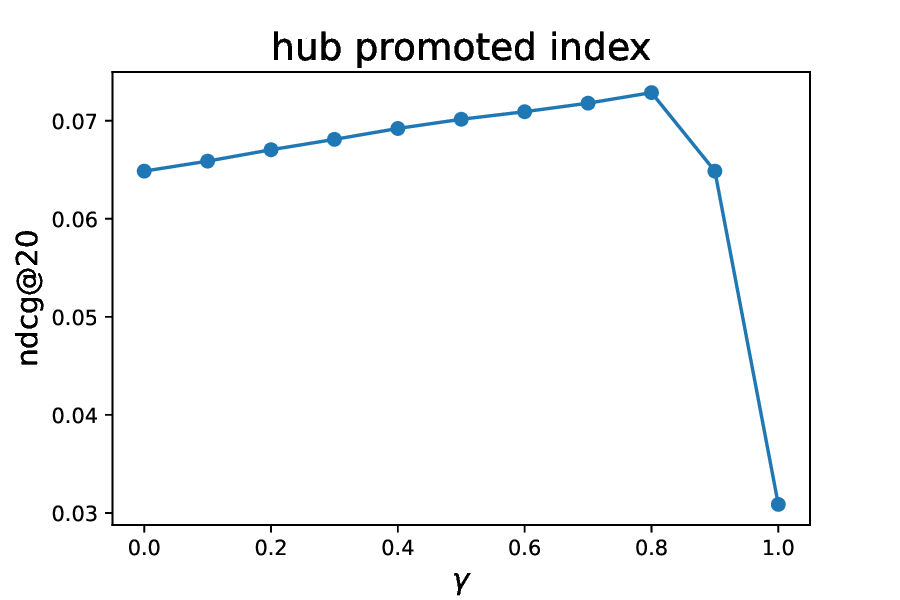}
	\includegraphics[width=0.3\textwidth]{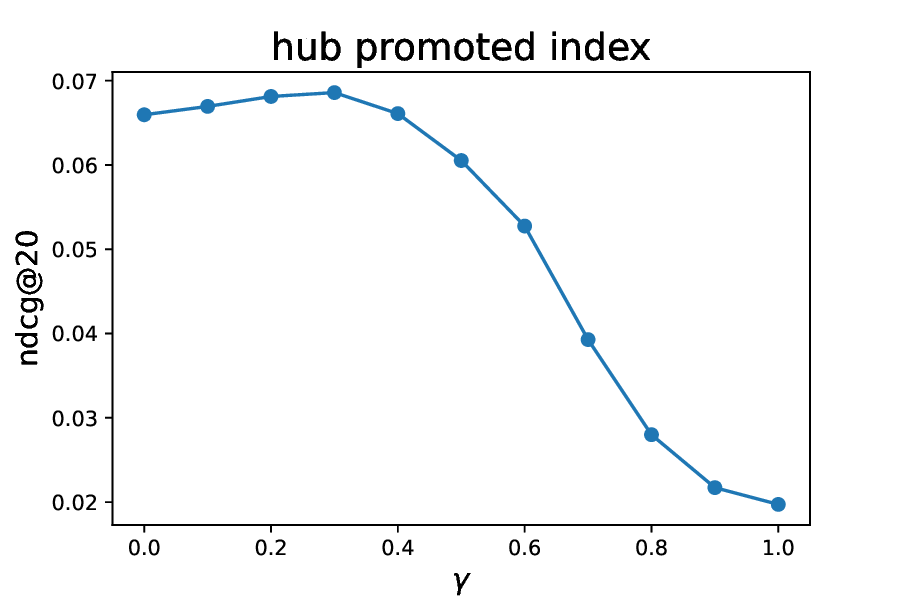}\\
	\includegraphics[width=0.3\textwidth]{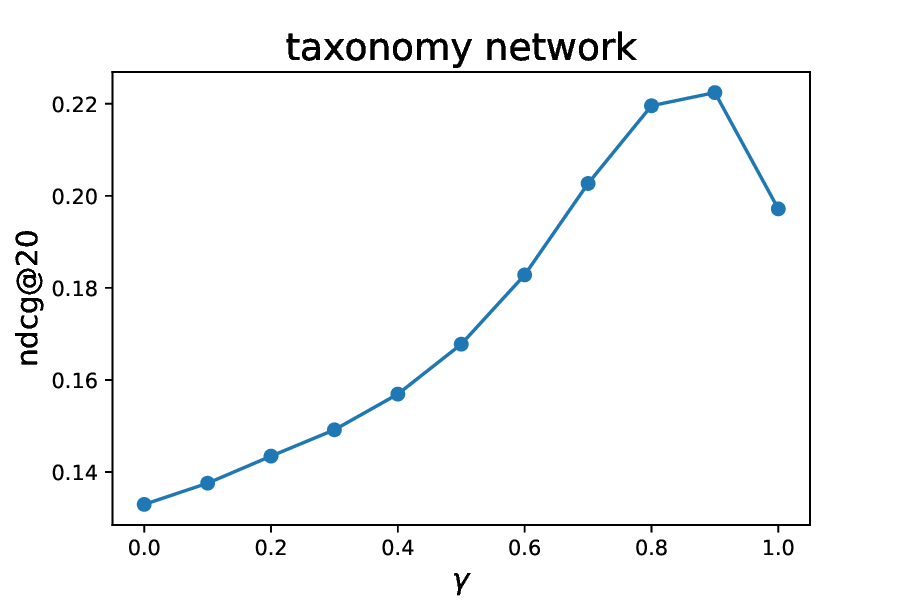}
	\includegraphics[width=0.3\textwidth]{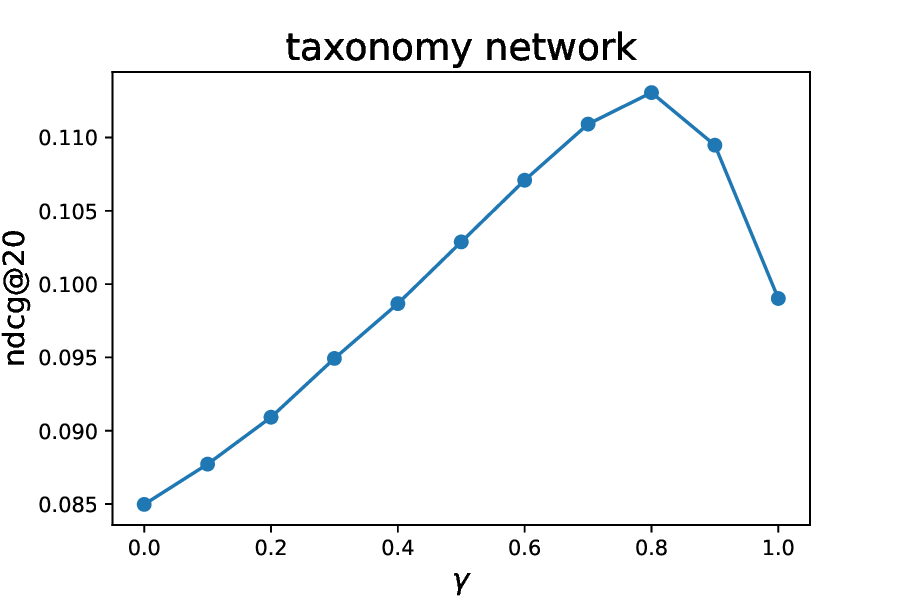}
	\includegraphics[width=0.3\textwidth]{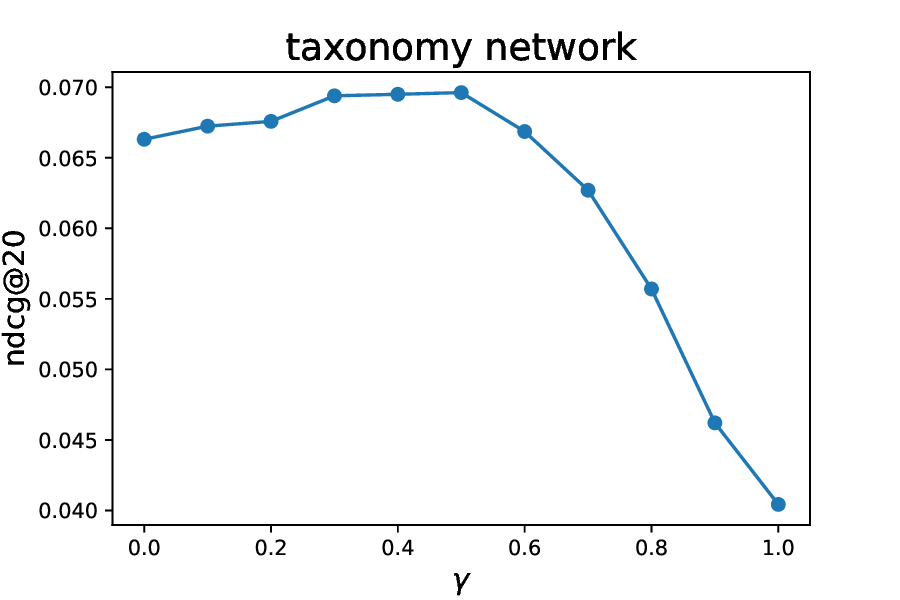}\\
	\includegraphics[width=0.3\textwidth]{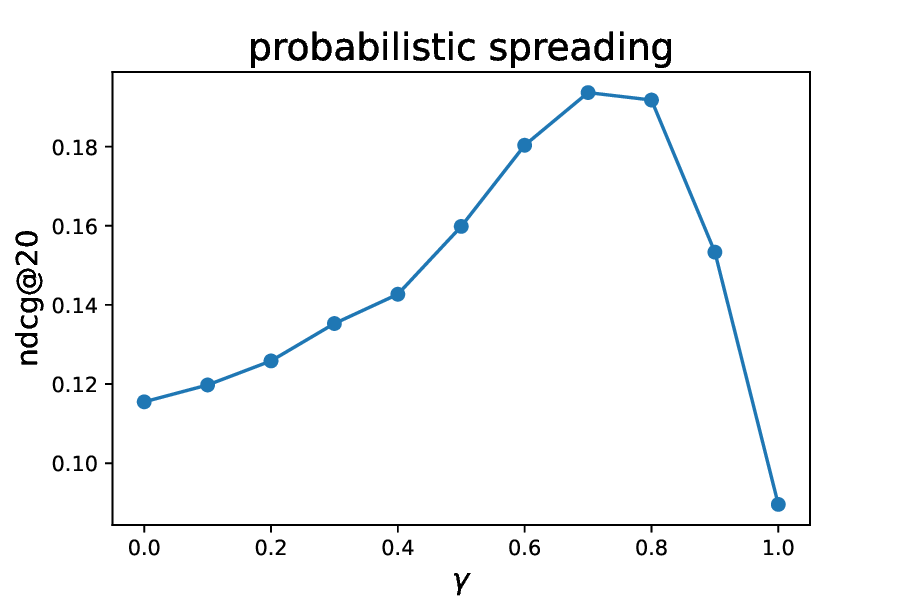}
	\includegraphics[width=0.3\textwidth]{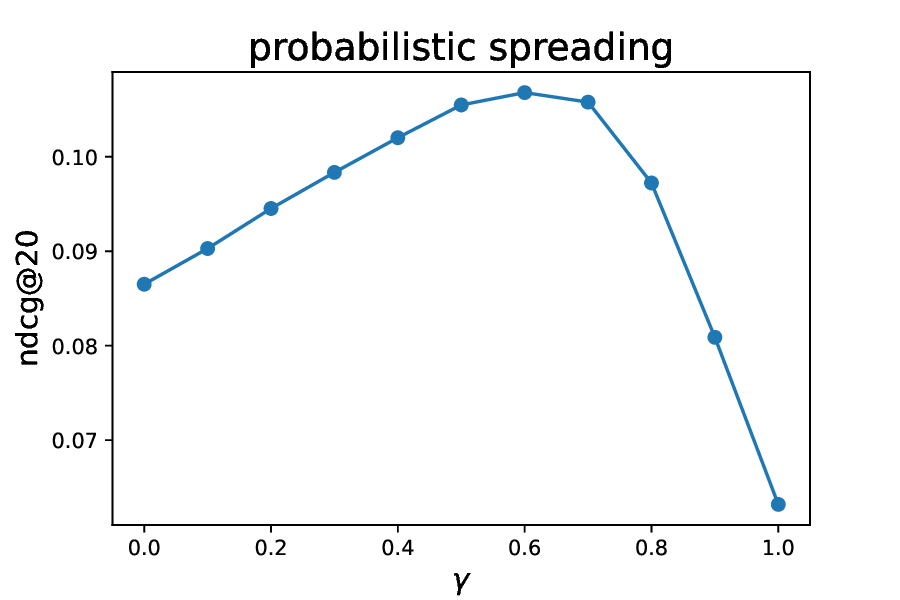}
	\includegraphics[width=0.3\textwidth]{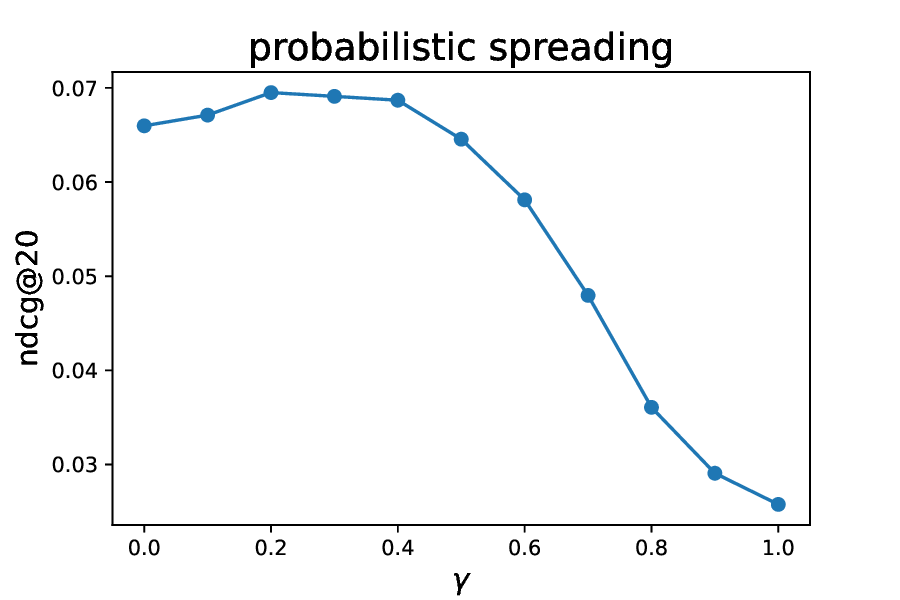}\\
	\includegraphics[width=0.3\textwidth]{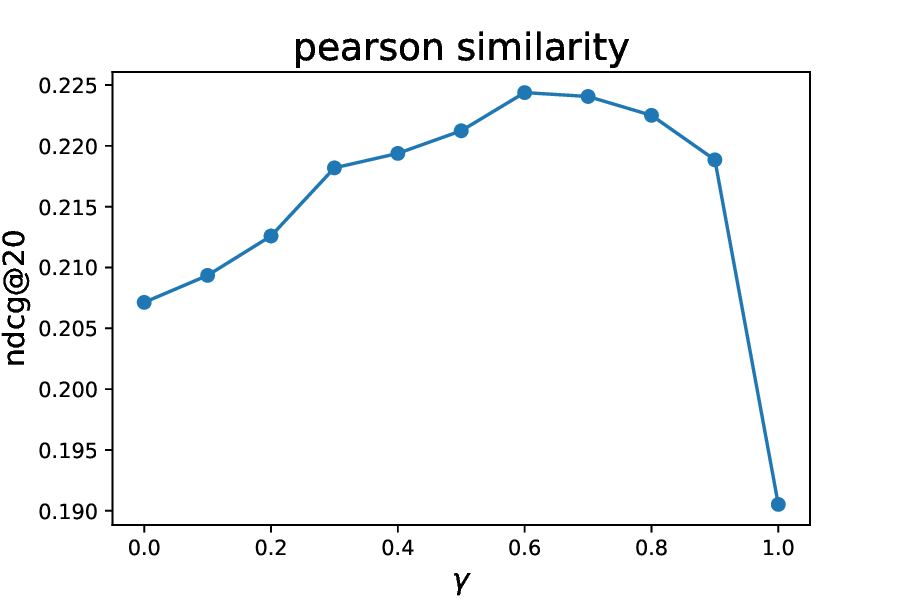}
	\includegraphics[width=0.3\textwidth]{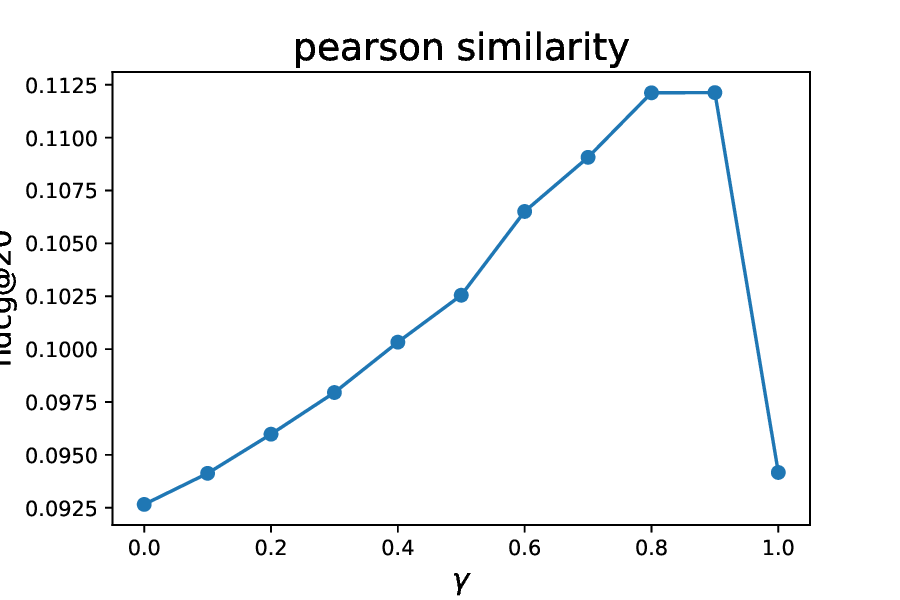}
	\includegraphics[width=0.3\textwidth]{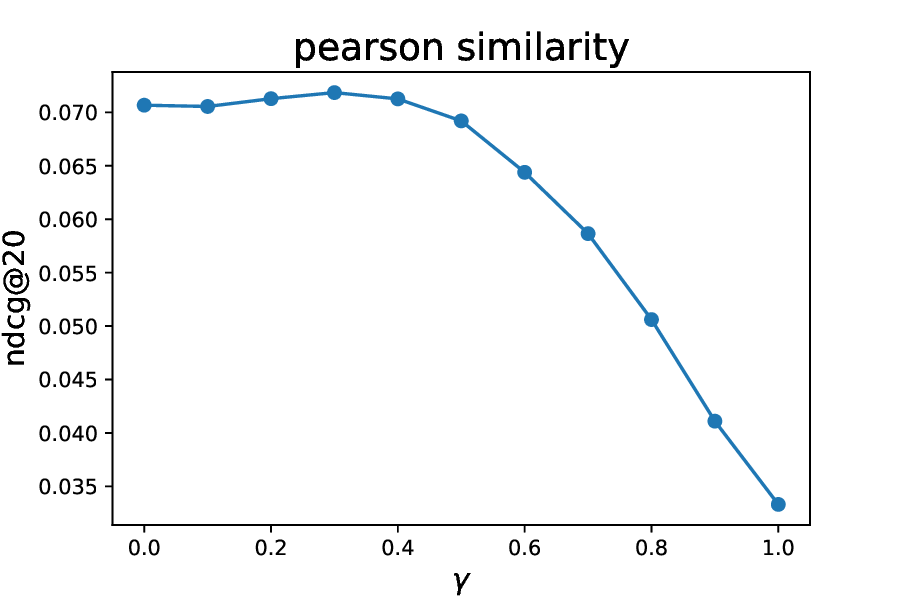}\\
	\includegraphics[width=0.3\textwidth]{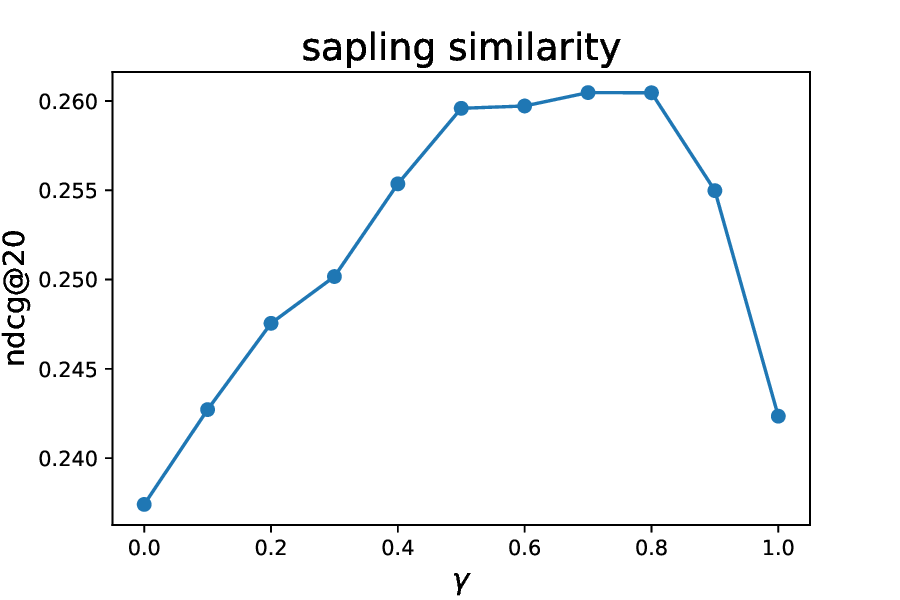}
	\includegraphics[width=0.3\textwidth]{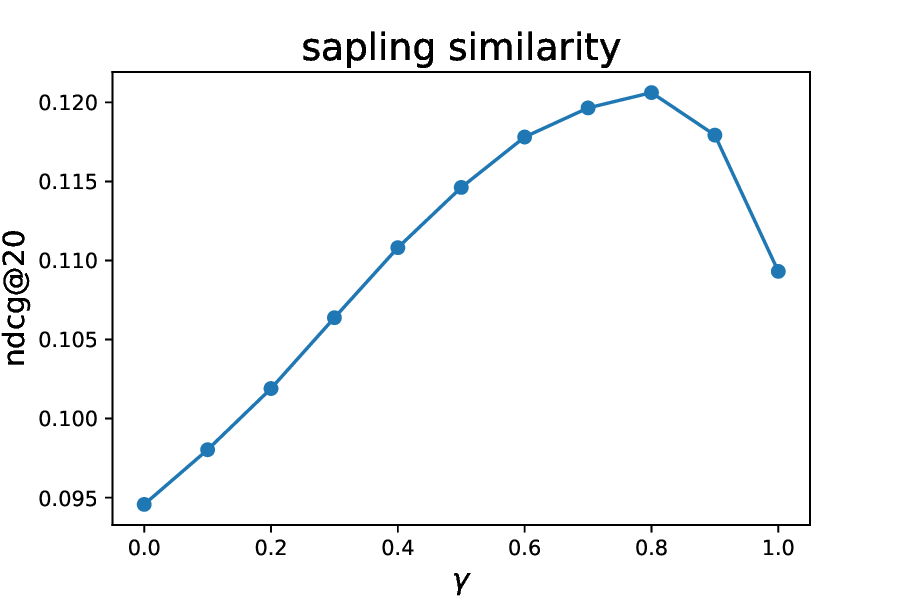}
	\includegraphics[width=0.3\textwidth]{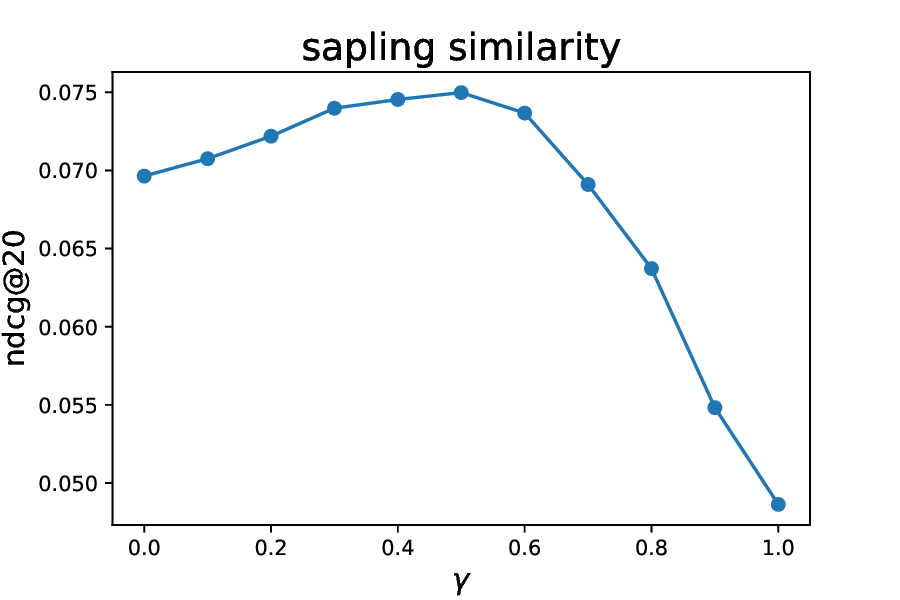}\\
	\caption{Optimization of the $\gamma$ parameter used to combine user-based and item-based approaches in country-export (first column), Amazon-product (second column), and Milan GPS data (third column).}
	\label{fig:gamma_small2}
\end{figure}

In figure \ref{fig:gamma_small1} and \ref{fig:gamma_small2} we show the optimization of the $\gamma$ parameter which determines the relative weight we assign to the user-based and the item-based collaborative filtering when computing the weighted average. We looked for the optimal value in the set \{0,0.1,0.2,0.3,0.4,0.5,0.6,0.7,0.8,0.9,1\}. $\gamma = 0$ is equivalent to the user-based approach and  $\gamma = 1$ is equivalent to the item-based approach, however, the ndcg@20 values are different from the ones in table \ref{tab:results_item_user} because to find the optimal value of $\gamma$ we use the validation set that, for each user, is composed by the 10\% rounded up of links in the train set.

\section*{S4. Identification of significant links in similarity networks}
\begin{figure*}[h!]
\centering
\includegraphics[width=1\textwidth]{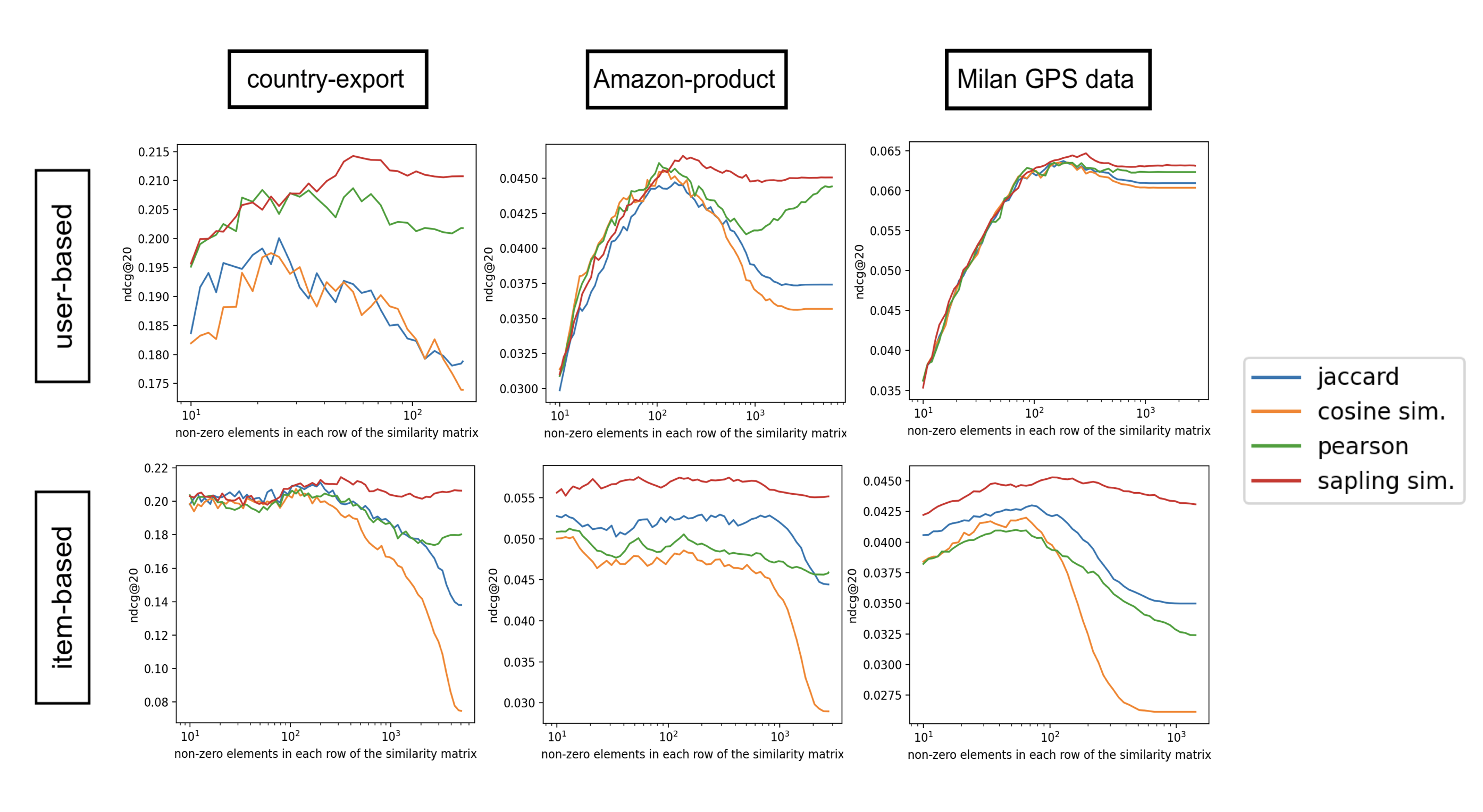}
\caption{The effect of setting elements in each row of the similarity matrix to zero, except for the k ones with the highest absolute values. We show the value of ndcg@20 for Jaccard, Cosine Similarity, Pearson and Sapling Similarity. The latter reaches the highest performance with all the datasets, moreover, we can see that it is less penalized by high values of k which means it is not affected by noisy relations among users or items.}
\label{fig:filtri}
\end{figure*}

Many of the similarity values between users or items are not statistically significant and must be considered noisy \cite{cimini2022meta}; When building a collaborative filtering they should be removed to avoid reducing the quality of recommendations.\\
For example, in the bipartite country-product network, most products share at least one co-occurrence with at least another product, and the same is true for countries. A single co-occurrence is sufficient for similarity metrics such as Jaccard and Cosine Similarity to have a non-zero (and positive) similarity. The purpose of this section is to investigate how identifying significant elements in the similarity matrix (as done, for example, by Pugliese et al. \cite{pugliese2019unfolding}, who used the Configuration Model as a null model \cite{saracco2017inferring}) affects the recommendations provided by similarity metrics.\\
Let us consider the item-based case, our approach to filter the similarity matrix consists of selecting for each item (row of the matrix) the k items (columns of the matrix) with the highest absolute value of similarity and setting the similarities with other items equal to zero. In the user-based case we follow the same procedure.\\
When computing the recommendation score of an item $\alpha$ to a user $i$ with a item-based Collaborative Filtering, we are involving in the computation only the k items that are really similar or, in the case of the Sapling Similarity and Pearson Similarity, also dissimilar to $\alpha$. By varying k, we look at how good are the recommendations provided in the three dataset: country-export, amazon-product and Milan GPS data.\\
In figure \ref{fig:filtri} we show how the value of k affects the recommendation quality (in terms of ndgc@20) of different similarity metrics both in the user-based and the item-based cases. As the reader can see Sapling Similarity generally performs better than the other metrics, moreover it is less penalized by using a high value of k meaning that it is less influenced by the presence of noisy relations between users or items.

\section*{S5. Sapling Similarity Network of countries}
\begin{figure*}[h!]
	\centering
	\includegraphics[width=1\textwidth]{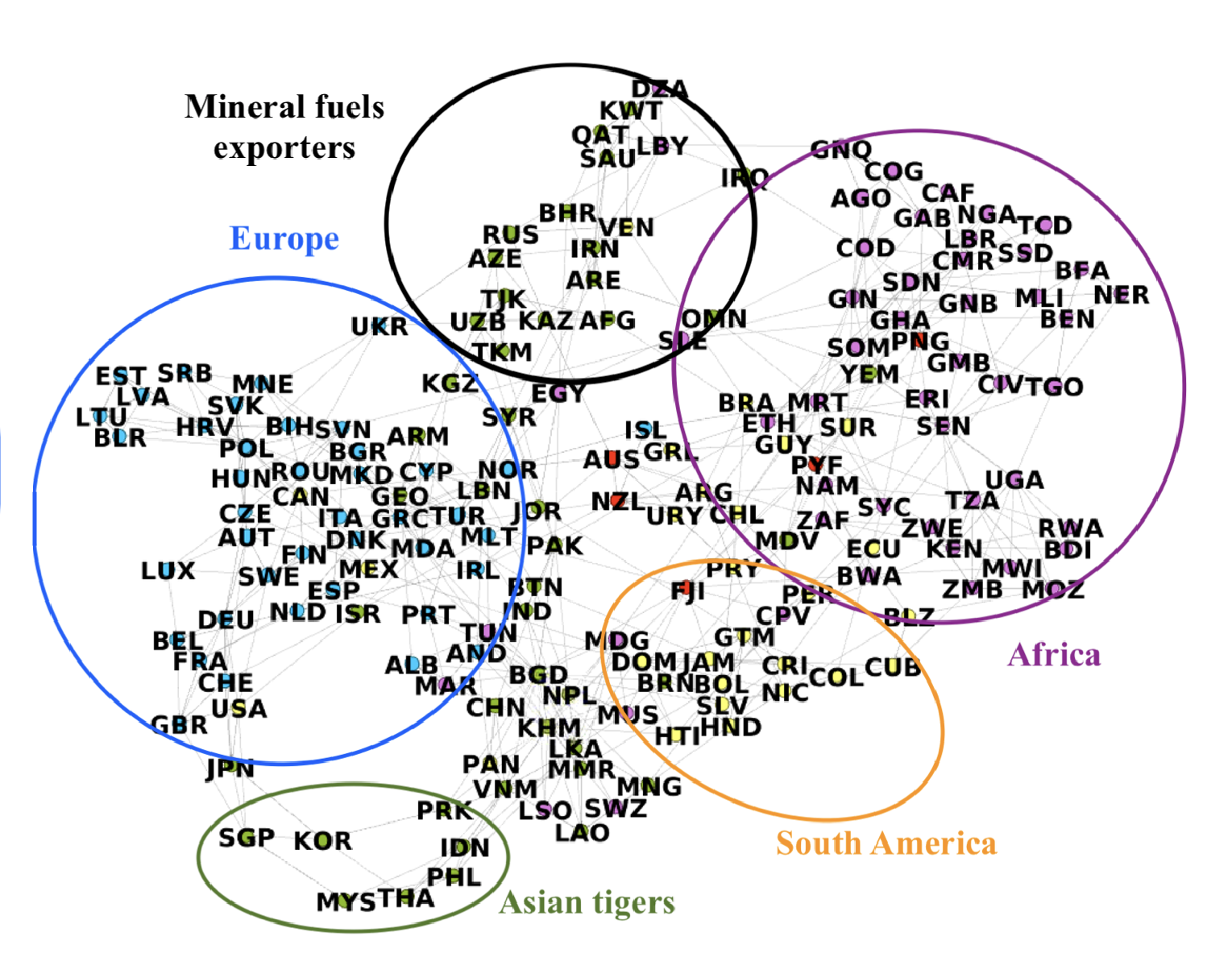}
	\caption{The Sapling Similarity Network (country layer). For each country only the top 4 links, in terms of the Sapling Similarity, are shown. The resulting structure reflects both geographical and industrial affinity among countries.}
	\label{fig:country_network}
\end{figure*}

In addition to the good performance of the Sapling Similarity when used to recommend new products to countries in collaborative filtering, here we want to provide a visible proof of its good functioning using it to extract a network of countries, or in other words to project the country-product bipartite network into the layer of the countries. In the Sapling Similarity Network each node is a country; for each country we create a link if the linked country is among the 4 highest values in terms of Sapling Similarity. We show the result in figure \ref{fig:country_network}.
One can easily identify geographical clusters corresponding to the Europe and Africa regions; the clear distinction between Asiatic countries focused on mineral fuels, like Russia and Arabia, and the Asian tigers, like Singapore, South Korea, Malaysia, Thailand etc. is far from trivial. It is also interesting to notice how Venezuela is separated from the other countries of South America and it is close to Asiatic countries related to mineral fuels.\\

\section*{S6. LightGCN: Hyper-parameter Settings}
\begin{figure*}[h!]
	\centering
		\includegraphics[width=0.8\textwidth]{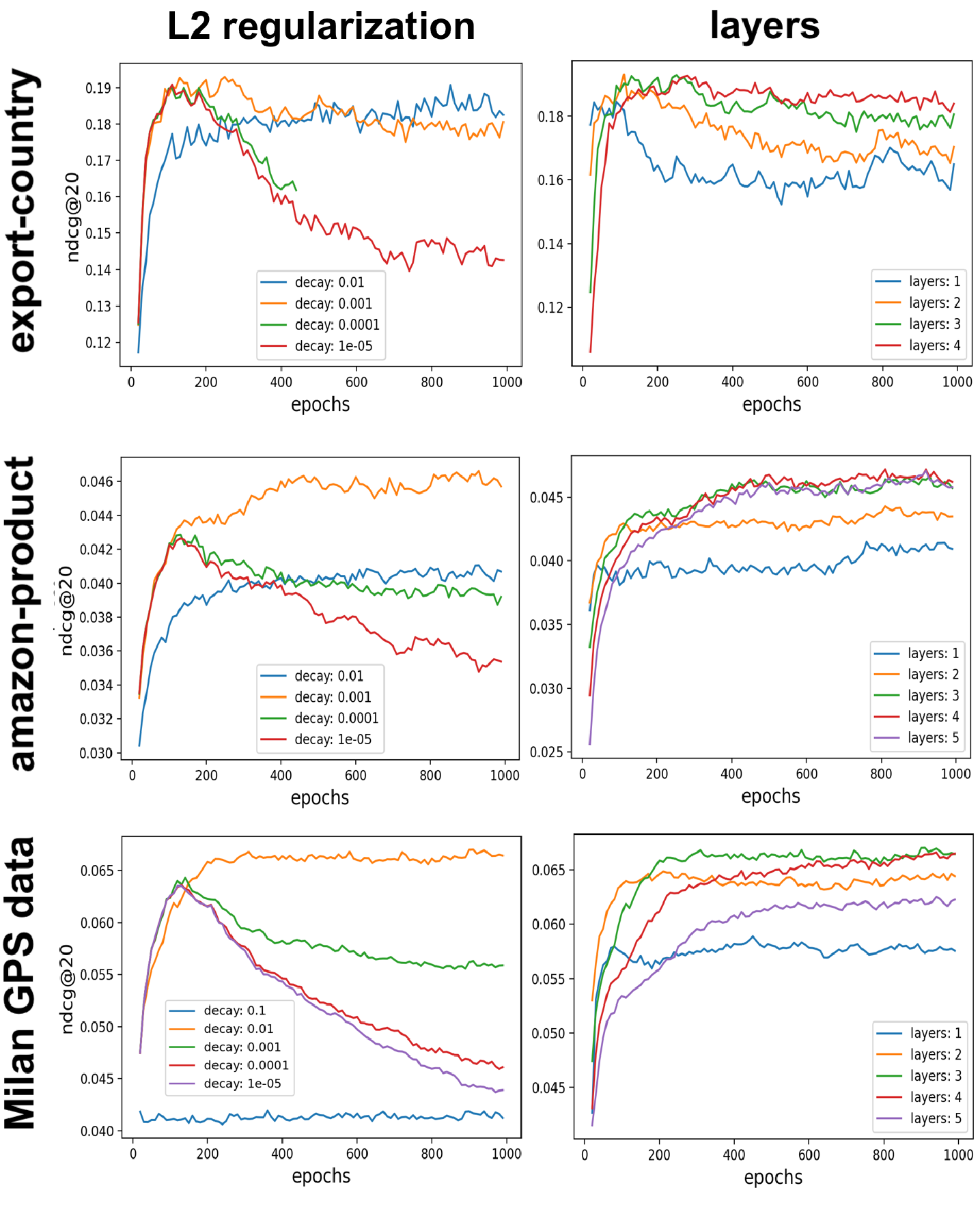}
		\caption{The optimization of L2 regularization term and the number of layers in the three dataset country-export, amazon-product, Milan GPS data.}
		\label{fig:LightGCN_tuning}
\end{figure*}

In this section we discuss the hyper-parameter settings for LightGCN applied to the three datasets country-export, amazon-product and Milan GPS data. Same as LightGCN authors \cite{he2020lightgcn} the embedding size is fixed to 64 and the optimizer is ADAM \cite{kingma2014adam} with the default learning rate 0.001 and default mini-batch size of 1024. We investigated the effect
of different values of the L2 regularization coefficient and the number of layers on the recommendations in the test set. In figure \ref{fig:LightGCN_tuning} we show the optimization of these two hyper-parameters.
We first tuned the L2 regularization coefficient and then the number of layers of LightGCN with the optimal value of the L2 regularization coefficient. The optimal values with which the results in the main paper are obtained are the following:
\begin{itemize}
\item \textbf{country-export:} L2 regularization = 1e-3, number of layers = 3, epoch = 130;
\item \textbf{amazon-product:} L2 regularization = 1e-3, number of layers = 4, epoch = 810;
\item \textbf{milan GPS data:} L2 regularization = 1e-2, number of layers = 3, epoch = 910;
\end{itemize}

\section*{S7. Non-Negative Matrix factorization: embedding size}
\begin{figure*}[h!]
	\centering
		\includegraphics[width=0.33\textwidth]{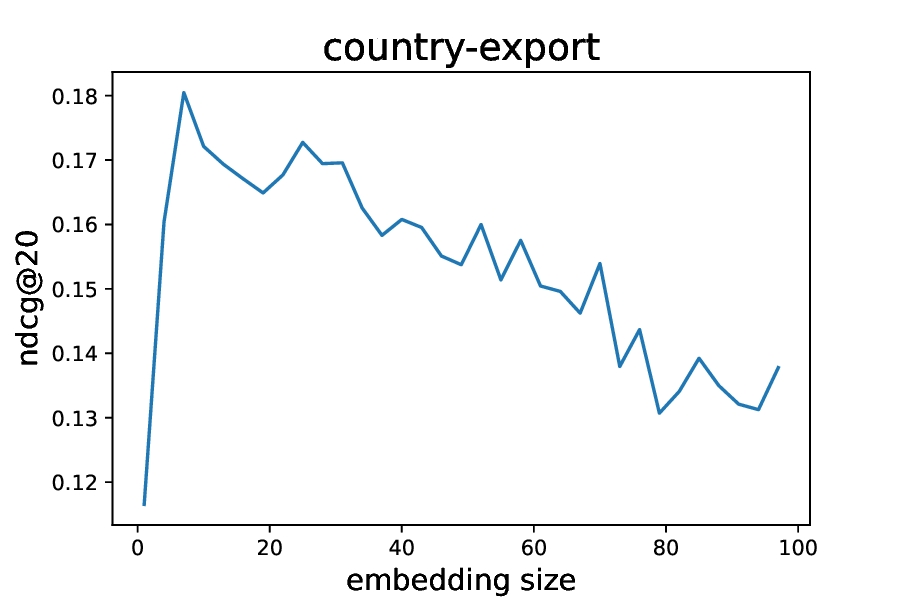}
            \includegraphics[width=0.33\textwidth]{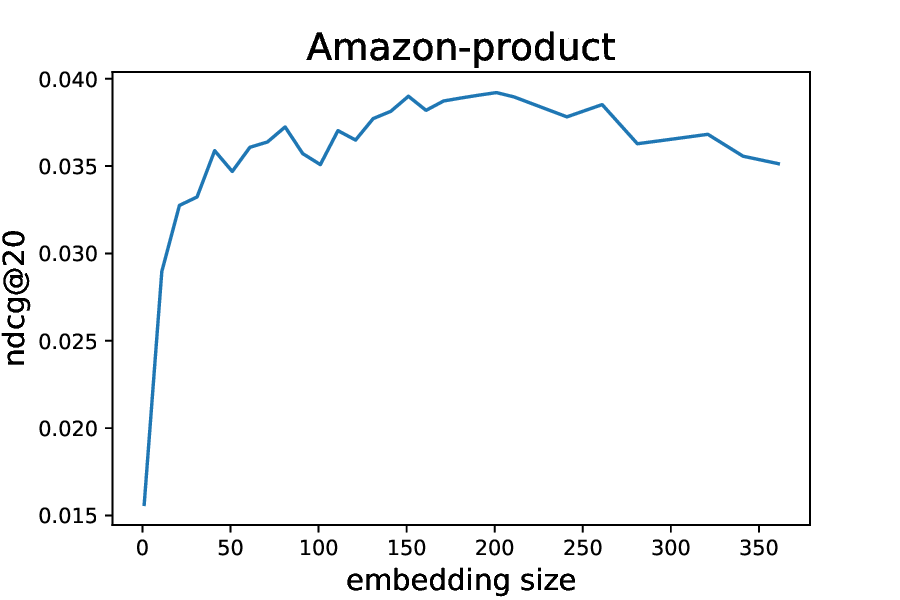}
            \includegraphics[width=0.33\textwidth]{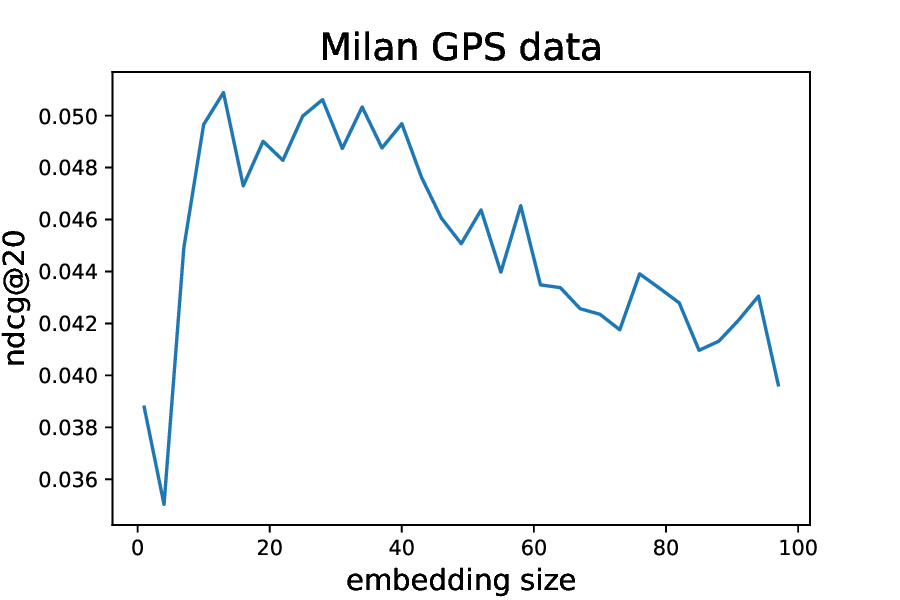}
		\caption{The effect of the embedding size to the quality of the recommendations provided by non-negative matrix factorization.}
		\label{fig:NMF_tuning}
\end{figure*}

In this section, we discuss the choice of the embedding size k used to apply non-negative matrix factorization. We investigated how the value of k affects the recommendation quality using the test set and we chose the optimal one. In particular, for the country-export dataset, we found k = 7, for the Amazon-product dataset we found k = 201 and for the Milan GPS dataset, we found k = 13. In figure \ref{fig:NMF_tuning} we show how the value of k affects the ndcg@20 score.
\section*{S8. Optimization of $\gamma$ in Gowalla. Yelp2018, and Amazon-book data}
\begin{figure*}[h!]
	\centering
		\includegraphics[width=0.33\textwidth]{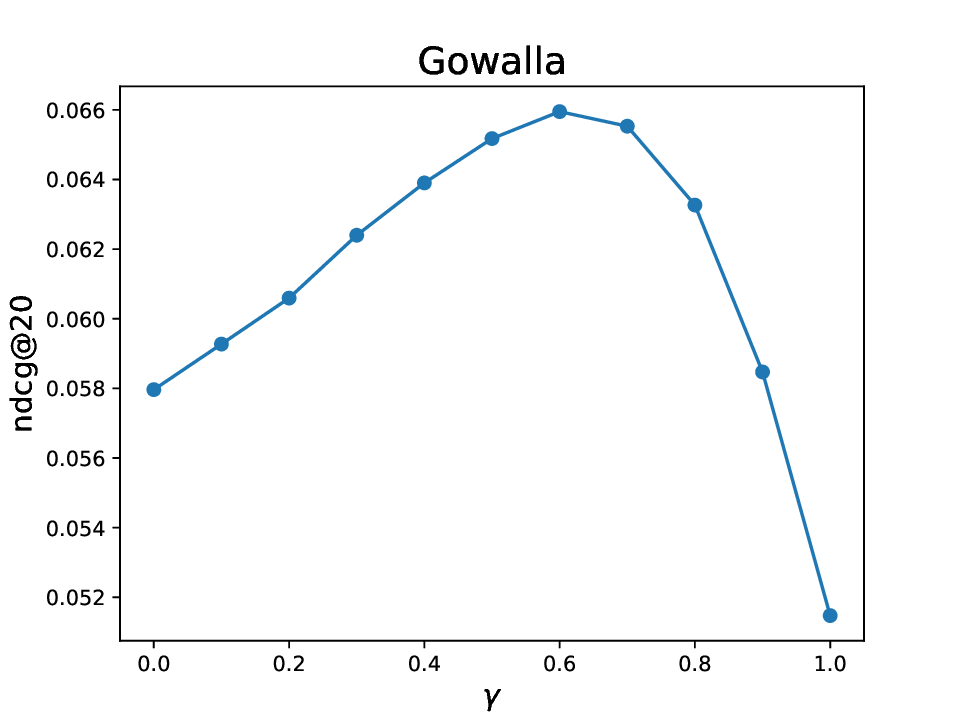}
            \includegraphics[width=0.33\textwidth]{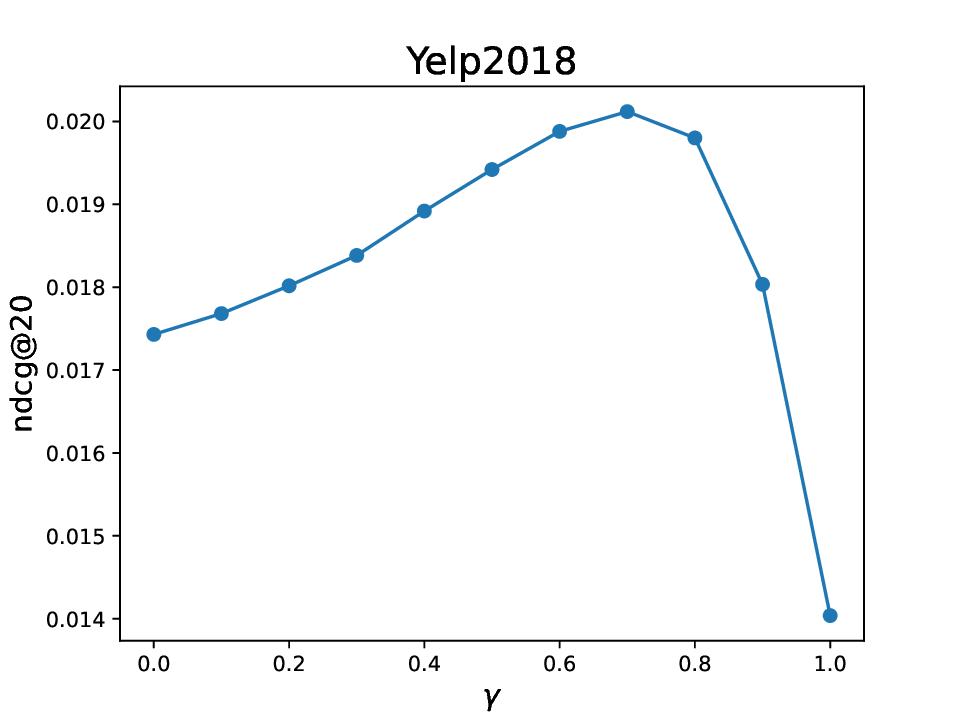}
            \includegraphics[width=0.33\textwidth]{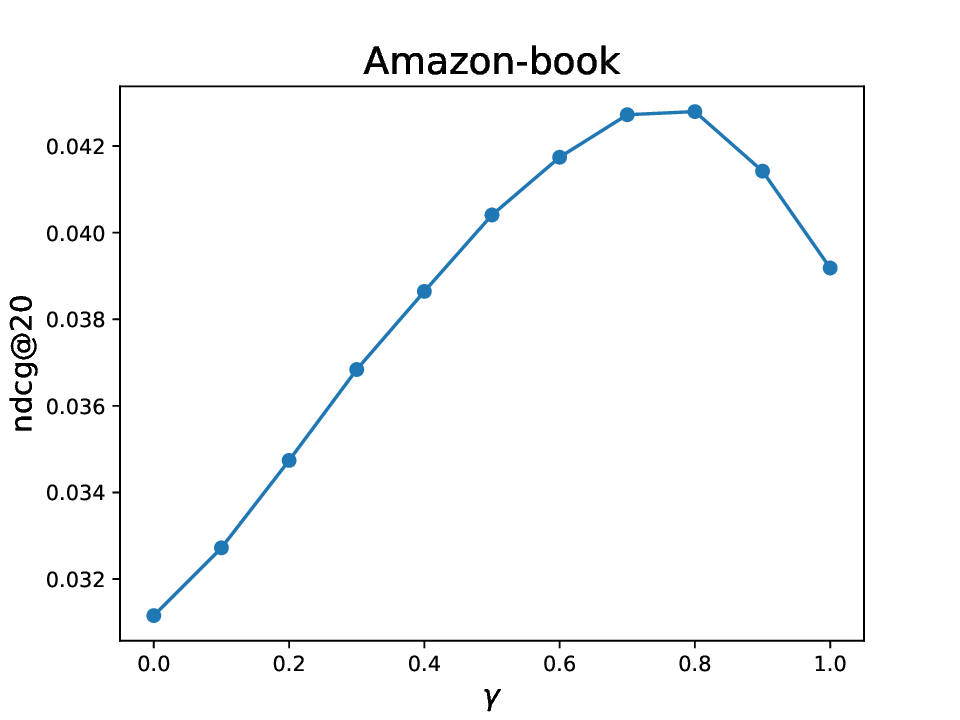}
		\caption{Optimization of the $\gamma$ parameter used to combine user-based and item-based approaches in Gowalla (left), Yelp2018 (center), and Amazon-book (right).}
		\label{fig:gamma_big}
\end{figure*}

In figure \ref{fig:gamma_big} we show the optimization on the validation set of the $\gamma$ parameter when building the SSCF model with Gowalla, Yelp2018, and Amazon-book data.

\section*{S9. Rating predictions on Movielens data}
In this section we compare similarities in predicting ratings users will give to movies. We use the movielens dataset (https://grouplens.org/datasets/movielens) that contains ratings users gave to movies in a time range of 1039 days. The ratings range from 1 to 5, and they are collected in a matrix $R$. As done with the Amazon-product dataset, to build the matrix $M$ we select the ratings equal to or above 3. To measure the similarity matrix $B$ we use a $M$ matrix built with data in the first 730 days. In this exercise we do not want to predict the simple presence of a link in the future, but we want to predict the rating a user $i$ will give to a movie $\alpha$. To compute our prediction we use these formulas:
\begin{equation}
S_{i\alpha}=\frac{\sum_{j\in W_\alpha}B^{(user)}_{ij}R_{j\alpha}}{\sum_{j\in W_\alpha}|B^{(user)}_{ij}|}~~~\text{ user based }
\label{rate_user}
\end{equation}
\begin{equation}
S_{i\alpha}=\frac{\sum_{\beta\in Q_i}B^{(item)}_{\alpha\beta}R_{i\beta}}{\sum_{\beta\in Q_i}|B^{(item)}_{\alpha\beta}|}~~~\text{ item based }
\label{rate_item}
\end{equation}
Where $W_\alpha$ is the set of users who rated the movie $\alpha$ in the first 730 days and $Q_i$ is the set of movies the user $i$ rated in the first 730 days (here we consider also ratings that are lower than 3). So, for instance, in the item-based case, our predicted rating for a user to a movie is the average rating the user gave to similar movies.\\
We then compare our predicted scores ($y^{pred}$) with the real scores ($y^{true}$) users gave to movies during the last 309 days using the following evaluation metrics:
\begin{itemize}
    \item MAE: the mean absolute error is $MAE = \frac{\sum_i|y^{true}_i-y^{pred}_i|}{N_r}$ with $N_r$ total number of ratings;
    \item RMSE: the root-mean-square error is $RMSE = \sqrt{\frac{\sum_i(y^{true}_i-y^{pred}_i)^2}{N_r}}$;
    \item ndcg: the normalized discounted cumulative gain is a measure of ranking quality that quantifies the goodness of the ranking of the recommendations. Details about this metric can be found in \cite{jarvelin2002cumulated}. Here we compute the ndcg separately for each user and then we compute the average.
\end{itemize}
In table \ref{tab:movielens} we show the results we get using different similarities. Also in this exercise, Sapling Similarity achieves the best scores.
\begin{table*}[ht]
\centering
\begin{tabular}{|l|ccc|ccc|}
\hline
&\multicolumn{3}{c|}{\textit{User-Based}}&\multicolumn{3}{c|}{\textit{Item-Based}}\\
\hline
& \textbf{MAE}   & \textbf{RMSE}     & \textbf{ndcg} & \textbf{MAE}   & \textbf{RMSE}    & \textbf{ndcg}\\ \hline
Common neighbors            & 0.730 & 1.010 & 0.907 & 0.822 & 1.139 & 0.901   \\ 
Jaccard                     & 0.730 & 1.010 & 0.909 & 0.781 & 1.090 & 0.906   \\ 
Adamic/adar                 & 0.729 & 1.010 & 0.908 & 0.824 & 1.141 & 0.901   \\ 
Resource allocation         & 0.728 & 1.009 & 0.909 & 0.830 & 1.150 & 0.900   \\ 
Cosine similarity           & 0.729 & 1.009 & 0.908 & 0.793 & 1.101 & 0.901   \\ 
Sorensen                    & 0.729 & 1.010 & 0.908 & 0.784 & 1.092 & 0.906   \\ 
Hub depressed index         & 0.729 & 1.010 & 0.908 & 0.776 & 1.083 & 0.904   \\ 
Hub promoted index          & 0.730 & 1.010 & 0.908 & 0.808 & 1.123 & 0.856   \\ 
Taxonomy network            & 0.728 & 1.009 & 0.908 & 0.773 & 1.081 & 0.906   \\ 
Probabilistic spreading     & 0.729 & 1.010 & 0.908 & 0.768 & 1.070 & 0.899   \\ 
Pearson Similarity          & 0.732 & 1.013 & 0.909 & 0.764 & 1.068 & 0.907   \\ 
Sapling Similarity          & \textbf{0.724} & \textbf{1.003} & \textbf{0.910} & \textbf{0.746} & \textbf{1.054} & \textbf{0.909} \\ 
\hline
\end{tabular}
\caption{Performance of the recommendation of user-based (on the left) and item-based (on the right) collaborative filtering that recommends new movies to users. the optimal model is the one with the lowest MAE and RMSE and highest ndcg. In both cases, Sapling Similarity outperforms the other similarities.}
\label{tab:movielens}
\end{table*}

\section*{S10. About the computational complexity of similarities}
\begin{figure*}[h!]
	\centering
		\includegraphics[width=0.7\textwidth]{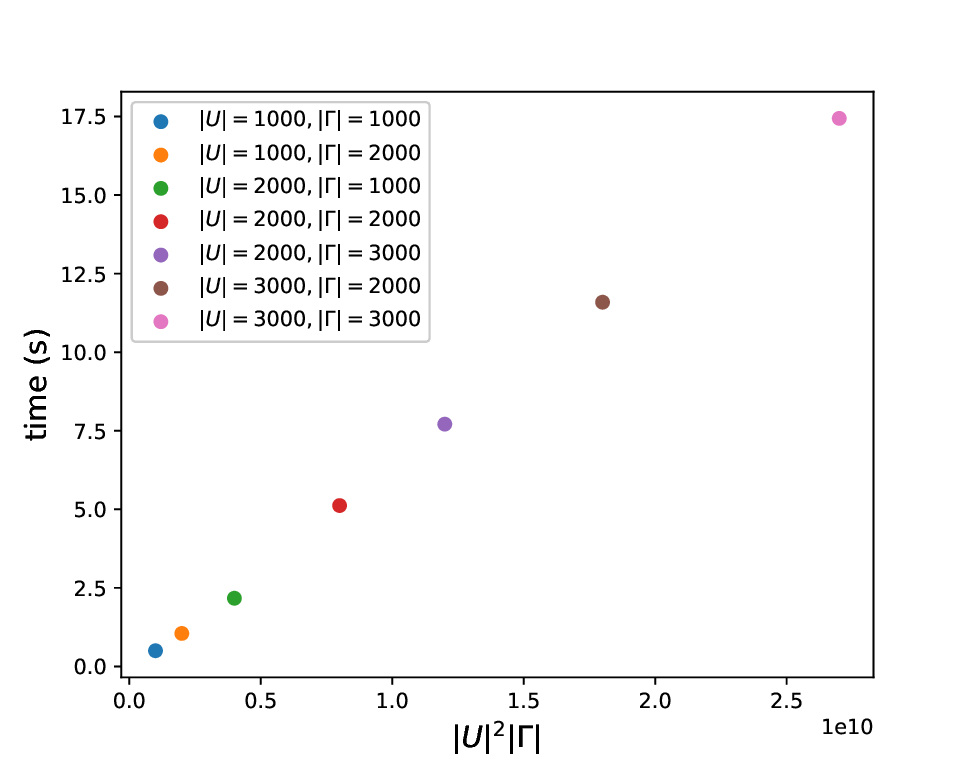}
		\caption{Computational time of Sapling Similarity in the user-based case on different bipartite networks. The density of links is fixed at $50\%$ and the values of $|U|$ and $|\Gamma|$ for each point are written in the legend. On the horizontal axis there is $|U|^2\times|\Gamma|$. As the reader can see from the linearity, the computational time of Sapling Similarity is $O(|U|^2|\Gamma|)$.}
		\label{fig:comp_comp}
\end{figure*}

All the similarity metrics we use in this study make use of the number of co-occurrences $CO_{ij}^{users}$ (or $CO_{ij}^{items}$) that can be computed with a matrix multiplication $MM^T$ (or $M^TM$). The standard computational time for a matrix multiplication $(n\times m)(m\times p)$ is $O(nmp)$ \cite{cormen2022introduction}, so in our case, being the dimension of $M$ $|U|\times|\Gamma|$ the computational time required for the computation of the number of co-occurrences is $O(|U|^2|\Gamma|)$ in the user-based case and $O(|U||\Gamma|^2)$ in the item-based case.\\
Regarding all the similarity metrics, the computational complexity can be reduced to the co-occurrence computation. We generated different random biadjacency matrices $M$ with a fixed density of links of $50\%$, varying the value of $|U|$ and $|\Gamma|$. In figure \ref{fig:comp_comp} we show the computational time of Sapling Similarity in the user-based case. As the reader can see the computational time is $O(|U|^2|\Gamma|)$.\\
We observe that there are several methods to compute the number of co-occurrences on a bipartite network, with different computational times. However, fixed the algorithm with which we compute the co-occurrences, all the similarity metrics we use in our study have the same computational complexity.
\end{document}